\documentclass[aoas]{imsart}

\RequirePackage{amsthm,amsmath,amsfonts,amssymb}
\RequirePackage[authoryear]{natbib}

\startlocaldefs

\usepackage{enumitem}
\usepackage{amssymb}
\usepackage{amsmath,amsfonts,bm}
\usepackage{adjustbox}
\usepackage{multirow}
\usepackage{amsthm}
\usepackage{microtype}
\usepackage{graphicx}
\usepackage{subfigure}
\usepackage{booktabs}

\usepackage{xcolor}
\newcommand{\E}{\mathbb{E}}

\usepackage[toc,page]{appendix}

\newtheorem{example}{Example}
\newtheorem{definition}{Definition}
\newtheorem{theorem}{Theorem}

\newtheorem{corollary}{Corollary}[theorem]

\newcommand{\uargmin}[1]{\operatorname{arg}\underset{#1}{\operatorname{min}}\;}
\newcommand{\RobustTop}{49.14}
\newcommand{\NaturalTop}{90.96}
\newcommand{\uargmax}[1]{\underset{#1}{\operatorname{arg}\,\operatorname{max}}\;}

\usepackage{hyperref}

\usepackage{algorithm}
\usepackage{algorithmic}
\usepackage[utf8]{inputenc}

\endlocaldefs

\begin{document}

\begin{frontmatter}
\title{Robust Sensible Adversarial Learning of Deep Neural Networks for Image Classification}
\runtitle{Robust Sensible Adversarial Learning}

\begin{aug}
\author[A]{\fnms{Jungeum} \snm{Kim}\ead[label=e1,mark]{kim2712@purdue.edu}}
\and
\author[A]{\fnms{Xiao} \snm{Wang}\ead[label=e2,mark]{wangxiao@purdue.edu}}
\address[A]{Department of Statistics, Purdue University, \printead{e1,e2}}

\end{aug}
\begin{abstract}

The idea of robustness is central and critical to modern statistical analysis. However, despite the recent advances of deep neural networks (DNNs), many studies
have shown that DNNs are vulnerable to adversarial attacks. Making imperceptible changes to an image can cause DNN models to make the wrong classification with high confidence, such as classifying a benign mole as a malignant tumor and a stop sign as a speed limit sign. 
The trade-off between robustness and standard accuracy is common for DNN models. In this paper, we introduce sensible adversarial learning and demonstrate the synergistic effect between pursuits of standard natural accuracy and robustness. Specifically, we define a sensible adversary which is useful for learning a robust model while keeping high natural accuracy. We theoretically establish that the Bayes classifier is the most robust multi-class classifier with the 0-1 loss under sensible adversarial learning. We propose a novel and efficient algorithm that trains a robust model using implicit loss truncation. We apply sensible adversarial learning for large-scale image classification to a handwritten digital image dataset called MNIST and an object recognition colored image dataset called CIFAR10. We have performed an extensive comparative study to compare our method with other competitive methods. Our experiments empirically demonstrate that our method is not sensitive to its hyperparameter and does not collapse even with a small model capacity while promoting robustness against various attacks and keeping high natural accuracy. The sensible adversarial learning software is available as a Python package at \url{https://github.com/JungeumKim/SENSE}.


\end{abstract}

\begin{keyword}
\kwd{Adversarial learning}\kwd{ Bayes classifier}\kwd{ classification}\kwd{ deep neural networks}\kwd{ implicit loss truncation}\kwd{ robustness}
\end{keyword}

\end{frontmatter}


\section{Introduction}

The idea of robustness is central and critical to modern statistical analysis, and it is about the idea that we can use models when we have realistic violations of model assumptions. \cite{tukey60} was the first to recognize the extreme sensitivity of some conventional statistical models to minor deviations from model assumptions. For Turkey's collective contributions to robustness, see \cite{huber02}, and for a historical review of robustness, see \cite{stigler10}. The theoretical aspects of robustness were presented in \cite{huber72}.

\begin{figure}
	\begin{center}
	\includegraphics[width=.95\linewidth]{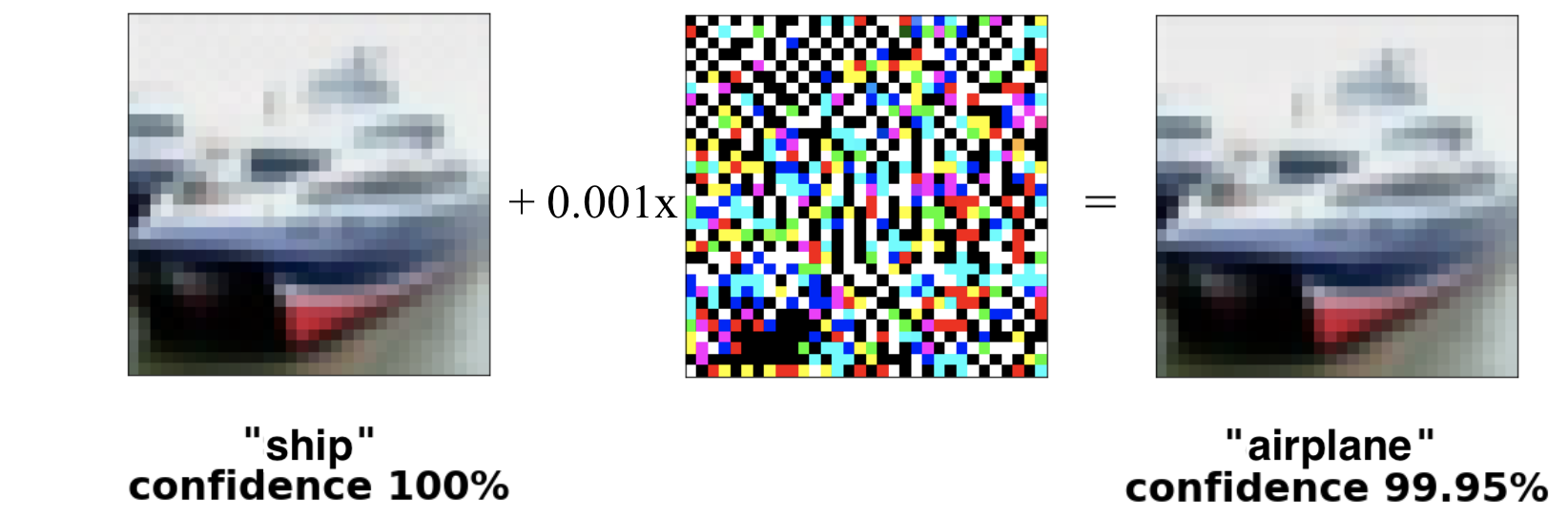}
	\end{center}
	\caption{A typical example of the vulnerablity of deep neural network classifiers. An imperceptibly small adversarial perturbation can significantly deviate the model prediction from a correct, confident prediction to an incorrect, yet still confident prediction. 
	}\label{fig:shipfool}
\end{figure}

In recent years, deep neural networks (DNNs) have demonstrated an amazing performance in 
solving many complex artificial intelligence tasks such as image classification and natural language processing \citep{goodfellow2016deep}. Given the prospect of a wide range of DNN applications, there are concerns regarding the safety and trustworthiness of DNNs.  
For example, many studies have shown that DNNs are vulnerable to adversarial attacks \citep{Dalvi2004Adversarial,biggio2013evasion, Tsipras2018There}. The adversarial inputs are called {\it adversarial examples}, typically generated by adding small perturbations that are imperceptible to human eyes \citep{Szegedy2013Intriguing}. Formally, if we take an example $x$ belonging to the class $c_1$ as input, there are efficient algorithms to find adversarial examples $x'$ such that $x'$ is very close to $x$, but the classifiers incorrectly predict it as belonging to class $c_2\neq c_1$. 
Many deep learning models achieve state-of-the-art (SOTA) performance in benchmark datasets, but they perform poorly on these adversarial examples. For example, a typical benchmark dataset for image classification is called CIFAR10, which consists of 60,000 $32 \times 32$ color images in 10 classes \citep{krizhevsky2009learning}.  \cite{Madry2017Towards} reported that the adversarial test accuracy on CIFAR10 is only 47\%; instead the natural test accuracy on CIFAR10 is around $95\%$ \citep{Springenberg2014Striving}. In Figure \ref{fig:shipfool}, a small amount of noise is added to an image of a ship in CIFAR10, which is classified by the network correctly with a $100\%$ confidence, but after adding this specially-crafted noise, this ship image is identified by the SOTA model as an airplane with a $99.95\%$ confidence.

Adversarial training (AT) is an empirical robust optimization to train a robust model to defend against adversarial attacks \citep{goodfellow2014explaining,kurakin2016adversarial,tramer2017ensemble,shafahi2019adversarial}. To deal with the min-max problem of the robust optimization, adversarial attacks typically are augmented with a perturbation constraint as the inner maximizers of the robust risk, and then the model is updated as an outer minimizer. This process is repeated until the robust risk converges. For example, the seminal work of AT by \cite{Madry2017Towards} trains with adversarial attacks generated by the projected gradient descent (PGD) method, which we will call regular AT (R-AT). R-AT is one of the most popular AT methods and has been known as the most effective algorithm among many AT algorithms \citep{carlini2019evaluating}. As an opposite concept to AT, natural training (NT) refers to any method that trains a model with no particular intention to defend, such as the prescribed SOTA classifiers.

A network trained by AT tends to have lower natural accuracy than those trained by NT. This trade-off is observed even with a small perturbation. For example, the SOTA robust classifiers on the CIFAR10 consistently show a clear trade-off even when the training perturbations are too small to be recognized by human eyes \citep{Madry2017Towards, zhang2019theoretically, tsipras2018robustness}. A vast volume of research efforts have been committed to understanding and resolving the trade-off problem; for example, from the perspective of sample complexity \citep{schmidt2018adversarially,yin2018rademacher,  stutz2019disentangling, lamb2019interpolated,raghunathan2019adversarial,carmon2019unlabeled,stanforth2019labels}, actual class change by adversarial perturbations \citep{tsipras2018robustness,suggala2018revisiting,stutz2019disentangling}, or model capacity of DNNs \citep{suggala2018revisiting,raghunathan2019adversarial}, or through alternative AT algorithms \citep{balaji2019instance,zhang2019theoretically, Wang2020Improving,Ding2020MMA}. These works are discussed with more details in the next section. At present, to the best of our knowledge, no single approach has predominantly been accepted. Moreover, many works that explain the trade-off often provide no algorithm to solve it or algorithms that require additional networks or datasets.

 In this paper, we propose a novel framework, called \emph{sensible adversarial robustness}, to relieve the trade-off between natural accuracy and robustness. In particular, we restrict adversarial perturbations not to cross the Bayes decision boundary besides the $\epsilon$-ball constraint so that the perturbation ball is adaptively modified for every single data point. Furthermore, we propose a practical algorithm to train a robust model under the proposed framework without any additional networks or auxiliary datasets. Theoretically, we provide a condition when no trade-off solution exists in terms of the data distribution support and the capacity of the model class. Algorithmically, when such a case happens, our proposed algorithm is identical to R-AT. When favorable conditions are not given, our sensible adversarial learning prioritizes natural accuracy to handle the trade-off. In addition, by considering the model capacity, we can practically deal with and explain the trade-off even with a small $\epsilon$ perturbation.

Our main contributions are:
(i). Under the framework of the sensible adversary, the pursuit of robustness and accuracy given enough model capacity can align with each other, i.e., there is no trade-off. We theoretically establish that the Bayes classifier is most robust against the sensible adversary. If the Bayes decision boundary can be far from data manifolds at least by $\epsilon$, our pursuit of sensible robustness does not cost any adversarial robust risk.
     (ii). We propose an efficient AT algorithm called SENSE-AT, which trains with sensible adversaries in the absence of the true Bayes classifier. It uses a novel technique called implicit loss truncation to handle true class change and limited model capacity. This SENSE-AT enjoys robustness without a significant drop in natural accuracy. Furthermore, the algorithm is not sensitive to the model capacity. When insufficient model capacity is given, our algorithm does not collapse to a constant function. Instead, it trains a model as robust as possible. 
    %
      %
      (iii). We experimentally demonstrate that SENSE-AT enables stably learning a robust and accurate model.  In particular, on CIFAR10, we achieve more than 90\% natural test accuracy and comparable adversarial test accuracy against various attacks. 
      

\section{Background and Related Work}
In this section, we review the adversarial learning framework and its trade-off between natural accuracy and adversarial robustness. 
\subsection{Adversarial learning}
The adversarial learning can be formulated as a minimization of the expected maximum loss within a fixed sized $\epsilon$-neighborhood of the data \citep{Madry2017Towards}.
Consider a $K$-class classification problem. Let $(X, Y)\in {\cal X}\times {\cal Y}$ be from an unknown distribution $\mathbb P_{X, Y}$. Let ${\cal F}$ denote the class of functions represented by DNNs. For a classifier $f\in {\cal F}$ and an $x\in {\cal X}$, the score function $f(x) = (f_1(x), \ldots, f_K(x))$ assigns score $f_i(x)$ to the $i$-th class. Let 
\begin{equation}\label{equ:hatpy}
\hat p_y (x) = {\exp(f_y(x)) \over \sum_{k=1}^K\exp(f_k(x))}.
\end{equation}
Then the predicted probability is $\hat p(x) = (\hat p_1(x),\ldots, \hat p_K(x))\in \cal P$, where $\cal P$ denotes the $(K-1)$-dimensional probability simplex. The predicted label of $x$ is $\hat y = \arg \max_i \{f_i(x)\}_{i=1}^K$. Let $\ell$ be the loss function  such as the 0-1 loss $\ell: {\cal Y}\times {\cal Y}\rightarrow \{0,1\}$ or the cross entropy loss $\ell: {\cal Y}\times {\cal P}\rightarrow \mathbb R$. In the later case, $\cal Y$ denotes the space of one-hot vectors, of which all elements are 0 except the 1 at the $y$-th dimension. Note that the standard natural learning is to find $f\in {\cal F}$ that minimizes the standard classification risk \[\mathcal{R}_{\rm{std}}(f)=\E\big[\ell(f(X), Y)\big].\] Unfortunately, its empirical minimization often does not yield models that are robust to adversarially crafted examples. In order to train a robust model, it is necessary to extend the natural learning framework appropriately. An $\epsilon$ perturbation ball is typically used to formally define the perturbations \emph{imperceptible} to human eyes. For each $x\in {\cal X}$, one can introduce a set of allowed perturbations and define the adversarial example  $x'$ such that $\|x' - x\|_p \le \epsilon$. This gives rise to the following robust adversarial classification risk with respect to  $\mathbb P_{X, Y}$ defined by
\begin{equation}\label{equ:robust_risk}
    \mathcal{R}_{\rm{rob}}(f) = \E\Big\{\max_{\|X'-X\|_p\le \epsilon} \ell(f(X'), Y)\Big\}.
\end{equation}
Formulations of this type have a long history in robust
statistics, going back to \cite{wald45}. 
\subsection{Adversarial examples and adversarial training}
Technically speaking, any perturbed example that is maliciously designed to decrease the classification performance of the classifier can be called an adversarial example. In the literature, however, the terminology \emph{adversarial example} is often used to indicate the inner maximizer in (\ref{equ:robust_risk}) as a stronger adversarial perturbation is more useful to test a trained DNN classifier's robustness. Formally, the adversarial example of $(x,y)$ w.r.t. $f$ can be defined as 
\begin{equation}\label{equ:inner_max}
 \tilde{x} = \arg\max_{\|x'-x\|_p\le \epsilon} \ell(f(x'), y).
 \end{equation}
 In this paper, we occasionally call $\tilde{x}$ as \emph{regular} adversarial example to clearly distinguish it from our proposed \emph{sensible} adversarial example. Among various methods to generate the adversarial examples, i.e., the inner maximizers, the most well-known methods are the Fast Gradient Sign Method (FGSM) \citep{good_explaining} and the projected gradient descent (PGD) \citep{Madry2017Towards}. FSGM proposes to compute the adversarial examples via the gradient of the loss on clean data $x$, i.e., 
\begin{equation*}
	\tilde{x}_{\rm FGSM} = x + \epsilon\cdot\mbox{sign}\big(\nabla_{x} \ell(f(x), y)\big).
\end{equation*}
This attack can be interpreted as a simple one-step scheme for generating (\ref{equ:inner_max}) under the $\ell_\infty$ norm.  PGD is another popular algorithm, which is the generalization of FSGM. PGD is a more powerful adversary with the multi-step variant. At the $t$-th step, the perturbed example is
\begin{equation}\label{equ:pgd}
		\tilde{x}_{\rm PGD}^{(t+1)} = \Pi_{x, \epsilon}\Big(	\tilde{x}_{\rm PGD}^{(t)} + \alpha\cdot\mbox{sign}\big(\nabla_{x} \ell(f(x), y)\big)\Big),
\end{equation}
where $\alpha$ is the learning rate and $\Pi_x$ is the projection operator which projects the data back to the set $\{x': \|x' - x\|_\infty\le \epsilon\}$. 

In practice, the risk in (\ref{equ:robust_risk}) is minimized by empirical risk minimization, and the loss $\ell$ is usually the cross-entropy loss. Especially, an empirical minimization approach that augments adversarial examples and uses them to evaluate the empirical risk is generally called AT. For example, the well-known R-AT algorithm generates the adversarial examples by PGD, and trains a robust model on them. Note that R-AT generally requires a larger capacity than NT \citep{gao2019convergence,xie2020intriguing}. If the model capacity is only sufficient for NT, R-AT may collapse to a constant function \citep{Madry2017Towards}. Despite the increased model capacity requirement, R-AT is known as the most effective method to train a robust DNN classifier \citep{carlini2019evaluating}.
\subsection{Trade-off in adversarial training} 
The trade-off problem in AT has been consistently reported and has sparked an enormous research interest. A hot research question was whether the trade-off is avoidable or inevitable. Many studies tried to answer the question from the perspective of sample complexity that R-AT requires more data for ensuring a good natural performance \citep{yin2018rademacher, schmidt2018adversarially, stutz2019disentangling, lamb2019interpolated,raghunathan2019adversarial,carmon2019unlabeled,stanforth2019labels}. If the trade-off were simply a problem of sample complexity, given infinite data, there were no trade-off. As an effort to increase the available sample size, \cite{carmon2019unlabeled} and \cite{stanforth2019labels} conducted self learning and significantly improved robustness. However, the resultant natural accuracy is not as high as that of natural learning (e.g., $<90\%$ for CIFAR10). Another stream of research that studied the trade-off focused on the actual class change by adversarial perturbations, which makes the trade-off inevitable. For example, when the maximum perturbation norm $\epsilon$ is large enough to change the true class, an inherent tension between pursuits of accuracy and robustness exists \citep{tsipras2018robustness}. Therefore, some works tried to prevent such class change by invalidating class-changing perturbations; \cite{suggala2018revisiting} theoretically showed that when only valid adversarial examples are considered, there is no trade-off. While \cite{suggala2018revisiting} provided no algorithm to realize their invalidation, \cite{stutz2019disentangling} realized the invalidation by utilizing autoencoders. Our work is built on their work, in that we also invalidate the class-changing adversarial example. Note that we provide a practical algorithm, and it does not require any additional networks other than the classifier itself. We also theoretically demonstrate no trade-off solutions under certain conditions, but the difference in the definition of adversarial perturbation makes our work orthogonal to the results in \cite{suggala2018revisiting}. Besides, we notice that the perspective of the actual class change does not explain the trade-off observed even for a smaller $\epsilon$ that does not cause any noticeable class change (e.g. CIFAR10 with $\epsilon=8/255$ in $\ell_\infty$-norm). To explain this, we also count model capacity as an important factor that causes the trade-off of R-AT with a small $\epsilon$. \cite{raghunathan2019adversarial} studied a trade-off when the model capacity is too rich and tried to solve it by using additional data to realize $n\rightarrow \infty$. Our work focuses on the another aspect of model capacity, where the capacity is not enough to achieve both high natural and robust accuracy.

As alternatives of R-AT to deal with the trade-off, a number of AT methods have been proposed \citep{balaji2019instance,zhang2019theoretically, Wang2020Improving,Ding2020MMA}. These methods do not require any additional network or auxiliary data to handle the trade-off. \cite{zhang2019theoretically} and \cite{Wang2020Improving} controlled the trade-off by a form of weight sum between natural and robust risk. \cite{balaji2019instance} optimized the perturbation limit $\epsilon$ for each input $x$, and \cite{Ding2020MMA} selectively applied a form of natural loss or robust loss. In Section 4, we will provide extensive comparative numerical studies between the proposed SENSE-AT and these methods in terms of both natural accuracy and adversarial accuracy. We also present methodological comparison in Appendix \ref{app:comp}.

\section{Sensible Adversarial Learning}

In this section, we propose a novel sensible adversary framework whose objective is to keep a low natural risk while decreasing adversarial risk.  We focus on a general multi-class classification problem with the 0-1 loss where $\mathcal{Y}=\{1, \ldots, K\}$. We consider $\ell_p$-norm constrained adversarial attacks, where $p\in\{0,1,...,\infty\}$.  

\subsection{Sensible adversary robustness}

We first give some definitions related to sensible adversarial robustness.  Let $(X, Y)\in {\cal X}\times {\cal Y}$ be from an unknown distribution $\mathbb P_{X, Y}$, and denote its Bayes classifier by $f^B$. Recall that the Bayes classifier would simply assign a test observation $x$ to the class $j$ for which $p_j(x)$ is largest, where $p_j(x)=\mathbb{P}(Y=j|X=x)$ with $j=1, \ldots, K$. 

 \begin{definition}\label{def:1}
{\rm Given a classifier $f\in {\cal F}$ with the function class ${\cal F}$, let $S_{x,\epsilon}(f) = \{z\in\mathcal{X}:\|z-x\|_p\leq \epsilon, f(z)=y\}$. 
The {\it sensible adversarial example} of $(x,y)$ w.r.t $f$ is defined as \begin{equation}\label{eq:sense_def}
\tilde{x}^s=\left\{
\begin{array}{ll}
x, & \textrm{if }f^B(x)\neq y,\\
\uargmax{z\in S_{x,\epsilon}(f^B)} \ell (f(z),y), & \textrm{if }f^B(x) =  y.
\end{array}
\right.
\end{equation}
}
\end{definition}

 Intuitively, the sensible adversarial examples would not cross the decision boundary of the Bayes classifier $f^B$, and this is achieved by adapting the perturbation ball for each example.
 In addition, a sensible adversary does not perturb a data point that the Bayes classifier incorrectly classifies. 

\begin{definition}
{\rm 
Let the {\it sensible adversarial loss} be $\ell^{s,B}_{\rm rob}(f,x,y)= \ell(f(\tilde{x}^s),y)$, where $\tilde{x}^s$ is the sensible adversarial example defined in Definition \ref{def:1}.
The {\it sensible robust risk} of a model $f$ is defined by \begin{equation}\label{eq:sense_loss}
    \mathcal{R}_{\rm rob}^{s,B}(f)=\E \Big[\ell^{s,B}_{\rm rob}(f,X,Y)\Big].
\end{equation}
We call its minimizer as a {\it sensibly robust model} w.r.t $\mathbb{\mathbb{P}}_{X,Y}$ and denote it by $f_{\rm rob}^{s,B}$ , i.e., $$f_{\rm rob}^{s,B} = \uargmin{f\in\mathcal{F}} \mathcal{R}_{\rm rob}^{s,B}(f).$$
}
\end{definition}

It is natural to expect that pursuing sensible robustness would \textit{not} cost natural accuracy because the attacks that are defended against cannot transgress the decision boundary of the Bayes rule. The following theorem confirms this intuition.  For a point $x$, denote the $\epsilon$-ball in $\ell_p$-norm centered at $x$ by $B(x,\epsilon)$. With a slight abuse of notation, for any set $A$, the $\epsilon$-neighborhood of $A$ is defined as $B(A,\epsilon)=\{y\in {\cal X}: \|y-x\|_p\leq \epsilon, x\in A\}$. 
Write the $\epsilon$-neighborhood of the decision boundary of a classifier $f$ as 
\[
DB(f, \epsilon) = \{x\in {\cal X}: ~\exists~ x' \in B(x,\epsilon) ~s.t.~f(x)\neq f(x')\}.
\]

\begin{theorem}\label{thm:1}
Let $\mathcal{R}_{\rm std}^*$ denote the minimum standard risk which is $\mathcal{R}_{\rm std}(f^B)$ with $f^B\in\mathcal{F}$, and $B(X,\epsilon)$ be an $\epsilon$-ball centered at $X$. Then
\begin{enumerate}
\item[(a).] $f^B$ is the unique minimizer of $\mathcal{R}_{\rm rob}^{s,B}(f)$ among $f\in\mathcal{F}$, i.e., $\mathcal{R}_{\rm rob}^{s,B}{(f^B)}= \mathcal{R}_{\rm std}(f^B)$.
\item[(b).] For any $f\in \mathcal{F}$ and any set $A\subset \mathcal{X}\setminus DB(f^B,\epsilon)$,
\begin{align*}
&\mathbb{P}\Big(\exists~ x' \in B(X,\epsilon)~ s.t.~ f^B(x')\neq Y, X\in A\Big) \leq \mathbb{P}\Big(\exists ~x' \in B(X,\epsilon) ~s.t.~f(x')\neq Y, X\in A\Big).    
\end{align*}
\end{enumerate}
\end{theorem}

According to Theorem \ref{thm:1}, a sensibly robust model w.r.t $\mathbb{\mathbb{P}}_{X,Y}$ is the Bayes classifier, i.e., $f^{s,B}_{\rm rob}=f^B$. This sensible robustness \textit{costs adversarial robustness} since $f^B$ may have a larger adversarial robust risk compared with $f_{\rm rob}$, a direct minimizer of $\mathcal{R}_{\rm{rob}}(f)$. However, Theorem \ref{thm:1} also shows that $f^B$ is equally or even more robust than $f_{\rm rob}$ except on $DB(f^B,\epsilon)$. Therefore, $f^B$ is most robust almost everywhere if the decision boundary of $f^B$ can lie outside of $B(\mathcal{X},\epsilon)$, so that $\mathcal{X}\setminus DB(f^B,\epsilon)=\mathcal{X}$. For example, this can happen when each class has its own support apart from each other by at least $2\epsilon$.

\begin{example}\label{ex:1}
{\rm 
We illustrate Theorem \ref{thm:1} with a toy example of a two-dimensional uniform distribution. Consider a two-dimensional random vector $X=(X_1,X_2)$ on $\mathcal{X}=(0,1)^2$ with a binary class $Y\sim Bernoulli(p)$, where $p>0.5$. Let the conditional distribution be 
\begin{align*}
    (X_1,X_2)\big|Y=0 & \sim Unif((0,\frac{1}{2})\times(0,1)),\\
    (X_1,X_2)\big|Y=1 &\sim Unif((\frac{1}{2},1)\times(0,1)).
\end{align*}
 By Theorem \ref{thm:1}(a), the optimal classifier w.r.t. the sensible adversarial loss is $f^B$. In this example, the Bayes classifier is $f^B(x)=sign(x_1-0.5)$, and $\mathcal{R}_{\rm{std}}(f^B)=0$ under the 0-1 loss. Meanwhile, when $\epsilon<1/4$, the robust classifier against $\epsilon$-ball attacks, which minimizes (\ref{equ:robust_risk}), is $f_{\rm{rob}}(x)=sign(x_1-(0.5-\epsilon))$. Its decision boundary is deviated by $-\epsilon$ from that of the Bayes rule, and this deviation costs natural accuracy by $\mathcal{R}_{\rm{std}}(f_{\rm{rob}})-\mathcal{R}_{\rm{std}}(f^B)=2 (1-p)\epsilon$. Note that as Theorem \ref{thm:1}(b) says, $f^B$ is most robust on any subset of $ \mathcal{X}\setminus DB(f^B,\epsilon)=(0, \frac{1}{2}-\epsilon)\times(0,1)\cup(\frac{1}{2}+\epsilon,1)\times(0,1)$. In fact, in this example, for any $x\in  \mathcal{X}\setminus DB(f^B,\epsilon)$, \[\max_{x'\in B(x,\epsilon)}\ell(f^B(x'),y)=0.\] We remark that the regular robust model $f_{\rm rob}$ is not robust on any point $x\in  (\max(0,\frac{1}{2}-2\epsilon), \frac{1}{2}-\epsilon)\times(0,1)\subset  \mathcal{X}\setminus DB(f^B,\epsilon)$, i.e., for such $x$,
\[\max_{x'\in B(x,\epsilon)}\ell(f_{\rm rob}(x'),y)=1.\] In this example, the complement of $\mathcal{X}\setminus DB(f^B,\epsilon)$ is not empty, and there is a trade-off of robustness for preserving high natural accuracy. The traded robustness compared with the regular $f_{\rm rob}$ is
$\mathcal{R}_{\rm{rob}}(f^B)-\mathcal{R}_{\rm{rob}}(f_{\rm{rob}})= \epsilon (4p-2).$ The derivation details are given in Appendix \ref{app:unif}. 
}
\end{example}

As Example \ref{ex:1} shows, our proposed framework regards the adversarial robust risk near the class border as \emph{reasonable gullibility}. According to this, we pursue robustness as long as such pursuit does not harm the natural accuracy on the data support. In other words, \emph{sensible robustness} respects the structure represented by data. 

When the data points reside in a high-dimensional space, the samples are too sparse to represent the true underlying distribution on the entire $\mathcal{X}\times \mathcal{Y}$. To take this phenomenon into account, 
consider a distribution $\tilde{\mathbb{P}}_{X,Y}$ on a subset $\tilde{\mathcal{X}}\times \mathcal{Y} \subset \mathcal{X}\times \mathcal{Y}$. 

\begin{theorem}\label{thm:2} Let $\mathcal{A}_\epsilon=\big\{f\in\mathcal{F}~\big|~ \tilde{\mathbb{P}}_{X,Y}\big(f(x)=f^B(x),~\forall x\in S_{X,\epsilon}(f^B)\big)=1\big\}$ and $\tilde{\mathcal{R}}_{\rm rob}^s(f) =\E_{\tilde{\mathbb{P}}_{X,Y}}[\ell^{s,B}_{\rm rob}(f(X),Y)]$. Then, for any $\epsilon>0$, $\tilde{\mathcal{R}}_{\rm rob}^s(f)$ is only minimized by any $f\in\mathcal{A}_\epsilon$. Furthermore, if $B(\tilde{\mathcal{X}},\epsilon)\supset {\mathcal{X}}, $ $f^B$ is the unique minimizer of $\tilde{\mathcal{R}}_{\rm rob}^s(f)$.
\end{theorem}

Theorem \ref{thm:2} provides a characterization of optimal classifiers w.r.t. the sensible adversarial loss when we only have data from $\tilde{\mathbb{P}}_{X,Y}$ restricted on $\tilde{\mathcal{X}}\times \mathcal{Y}$. It also provides a condition to find $f^s_{\rm rob}$, the optimal function w.r.t. $\mathbb{\mathbb{P}}_{X,Y}$. The benefit of the characterization of Theorem \ref{thm:2} is in the following Corollary \ref{co:1}.

\begin{corollary}\label{co:1}
Let $\mathcal{A}_0=\uargmin{f\in\mathcal{F}}\E_{\tilde{\mathbb{P}}_{X,Y}}[\ell_{\rm std}(f(X),Y)],$ which is the set of Bayes rules w.r.t. $\tilde{\mathbb{P}}_{X,Y}$.
Then, for any minimizer $f^*$ of $\tilde{\mathcal{R}}_{\rm rob}^s(f)$,
\[\mathcal{R}_{\rm std}(f^*) \leq \max_{f\in \mathcal{A}_0} \mathcal{R}_{\rm std}(f).\]
\end{corollary}

 Corollary \ref{co:1} shows that there is a synergistic effect between the pursuit of sensible robustness and natural accuracy. That is, the pursuit of sensible robustness rather promotes more natural accuracy. Now we illustrate these theoretical results in the following Example \ref{ex:2}.

\begin{example}\label{ex:2}
{\rm
   Consider the same setting and notation as in Example \ref{ex:1}. Assume that we only observe data 
generated from $\tilde{\mathbb{P}}_{X,Y}$ on $\tilde{\mathcal{X}}\times \mathcal{Y} \subset \mathcal{X}\times \mathcal{Y}$, and the training and test sets do not provide any information on $\tilde{\mathcal{X}}^c\times\mathcal{Y}$.
Assume the support is $\tilde{\mathcal{X}}=\cup_{i=1}^{3}\cup_{j=1}^{3}H_{ij},$ where $H_{ij}=((\frac{\alpha}{2},\frac{3\alpha}{2})+2\alpha(i-1))\times ((\frac{\alpha}{2},\frac{3\alpha}{2})+2\alpha(j-1))$, $\alpha = \frac{1}{6}$, and $j=1,2,3$, as illustrated in Figure \ref{fig:CheeseHoles}(a). Among all classifiers which predict the same as $f^B$ on $\tilde{\mathcal{X}}$, the worst one, denoted by $\tilde{f}^{B*}$,  predicts exact opposite on $x\in \tilde{\mathcal{X}}^c$ as illustrated in Figure \ref{fig:CheeseHoles}(b). For this worst case, the true standard risk w.r.t ${\mathbb{P}}_{X,Y}$ is $3/4$, although the standard risk w.r.t. $\tilde{\mathbb{P}}_{X,Y}$ is zero, i.e., 
\[
\mathcal{R}_{\rm std}(\tilde{f}^{B*})=3/4  \mbox{~~~and~~~}  \tilde{\mathcal{R}}_{\rm std}(\tilde{f}^{B*})=0. 
\]

Among minimizers of $\tilde{\mathcal{R}}_{\rm{rob}}$, let $\tilde{f}^*_{\rm{rob}}$ be the worst classifier in that it predicts incorrectly outside $B(\tilde{\mathcal{X}},\epsilon)$.  Specifically, let $\epsilon\leq \alpha/4$. Then $\tilde{f}^*_{\rm{rob}}$ classifies as depicted in Figure \ref{fig:CheeseHoles}(c). Compared with $\tilde{f}^{B*}$ on $B(\tilde{\mathcal{X}},\epsilon)$, $\tilde{f}^*_{\rm rob}$ should correctly classify, except on 
$$A_\epsilon = \{(x_1,x_2):0.5-\epsilon \leq x_1 \leq 0.5\}\cap B(\tilde{\mathcal{X}},\epsilon).$$ However, this increased accuracy of $\tilde{f}^*_{\rm{rob}}$ on $B(\tilde{\mathcal{X}},\epsilon)\setminus A_\epsilon$ is canceled out by its inaccuracy on $A_\epsilon$. Note that this inaccuracy is caused by the robustness on $\tilde{\mathcal{X}}$. Therefore, in this example, the regular robust learning does not have the synergistic effect of the sensible robust learning.

Now consider the worst-case sensibly robust classifier $\tilde{f}^{s*}_{\rm rob}$, which is illustrated in Figure \ref{fig:CheeseHoles}(d). Its decision boundary should not deviate to the left on $B(\tilde{\mathcal{X}}, \epsilon)$ no matter how large $\epsilon$ is. Its standard risks w.r.t. ${\mathbb{P}}_{X,Y}$ and $\tilde{\mathbb{P}}_{X,Y}$ are 
\[
\mathcal{R}_{\rm std}(\tilde{f}^{s*}_{\rm rob})=\max\big(0,{3\over 4}-9(4\epsilon^2+\frac{2}{3}\epsilon)\big) \mbox{~~~and~~~} \tilde{\mathcal{R}}_{\rm std}(\tilde{f}^{s*}_{\rm rob})=0.
\]
\begin{figure}[t]
	\begin{center}
	\includegraphics[width=1.\linewidth]{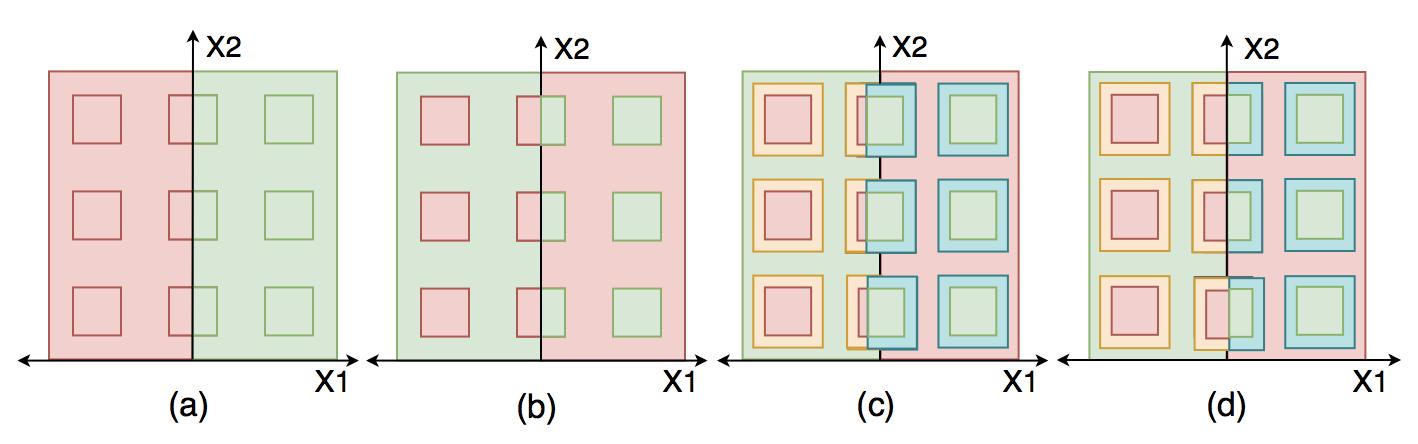}
	\end{center}
	\vspace{-0.4cm}
	\caption{Different classifiers in Example 2. (a) Supports ${\cal X}$ and $\tilde{\mathcal{X}}$. (b) The worst \textit{naturally} optimal model w.r.t. $\tilde{\mathbb{P}}_{X,Y}$. (c) The worst \textit{robustly} optimal model w.r.t. $\tilde{\mathbb{P}}_{X,Y}$. (d) The worst \textit{sensibly} robust model w.r.t. $\tilde{\mathbb{P}}_{X,Y}$.
	}	
	\label{fig:CheeseHoles}
	\vspace{-0.2cm}
\end{figure}

Therefore, $\tilde{f}^{s*}_{\rm rob}$ has a much more improved true standard risk than $\tilde{f}^{B*}$ although this benefit is not seen with the restricted data from $\tilde{\mathbb{P}}_{X,Y}$. Note that as $\epsilon$ increases, the true standard risk becomes zero, that is, $\tilde{f}^{s*}_{\rm rob}=f^B$.
}
\end{example}


\section{Sensible Adversarial Training} 


A transition from theory to the algorithm for actual training poses two main challenges. First, the 0-1 loss function is hard to optimize. The second challenge is that $f^B$ is not given. Rather, obtaining this Bayes classifier is our learning goal. In this section, we first present our remedy for the challenges, along with our training algorithm. Then, we discuss why this remedy solves the problem of non-availability of a true Bayes rule. 

\subsection{The objective function and algorithm}
For the first challenge, as a common practice, we replace it with the cross-entropy loss. Recall that $p_y(x)=\mathbb{P}(Y=y|X=x)$, and $\hat p_{y}(x)$ is the predicted probability of the label of $x$ being $y$ by a model $f$ in (\ref{equ:hatpy}). Then the cross-entropy loss is given by $\ell(f(x),y) = -\log \hat p_y(x)$ for $\hat p(x)\in \cal P$. In accordance with the loss change, we substitute the condition $f^B(x)=y$ by $\hat p_y(x)\geq c$ and $f^B(x)\neq y$ by $\hat p_y(x)< c$ for $c\in[0.5,1]$. Second, we extend the sensible adversarial risk in (\ref{eq:sense_loss}) to refer not to $f^B$ but the current function $f$ itself. Then the modification results in a training version of the sensible adversarial attack.

 \begin{definition}
 \label{def:3} 
 {\rm Given a classifier $f$, let $S_{x,\epsilon}(f) = \{z\in\mathcal{X}:\|z-x\|_p\leq \epsilon, \hat p_y(z)>c\}$. The sensible adversarial {\it training} example of $(x,y)$ w.r.t $f$ is defined as 
 \begin{equation}\label{eq:senseEg1}
\tilde{x}^s = \uargmax{x'\in \{x\}\cup S_{x,\epsilon}(f)} \ell (f(x'),y).
\end{equation}
}
\end{definition} From what follows, we use $\tilde{x}^s$ to denote the perturbed example defined in Definition \ref{def:3}, not in Definition \ref{def:1}. The sensible adversarial loss and risk also are redefined accordingly. 
\begin{definition}\label{def:4}
{\rm 
Let the sensible adversarial {\it training} loss be 
\begin{equation}\label{eq:sense_loss2}
\ell^{s}(f,x,y)= \ell(f(\tilde{x}^s),y)
\end{equation} where $\tilde{x}^s$ is a sensible adversarial training example as defined in Definition \ref{def:3}. The {\it training} sensible robust risk of a model $f$ is defined by \begin{equation}\label{eq:sense_risk2}
   \mathcal{R}^{s}(f)=\E \Big[\ell^{s}(f,X,Y)\Big].
\end{equation}}
\end{definition}
Now we introduce sensible adversarial training (SENSE-AT), an empirical minimization of (\ref{eq:sense_risk2}) with data augmentation. We choose $\tilde{x}^s$ in a way similar to the projected gradient descent (PGD) method of \citep{Madry2017Towards}, which is the standard method for large-scale constrained optimization. Comparing the optimization problem of $\tilde{x}$ in (\ref{equ:inner_max}) with that of $\tilde{x}^s$ in (\ref{eq:senseEg1}), we see that the later has an additional constraint that $ \hat p_y(\tilde{x}^s)>c$ i.e., $\ell(f(\tilde{x}),y)<-\log c$. Therefore, during the K-step of PGD iterations, once the loss of a currently generated example exceeds $-\log c$, we reverse it back to the previous step and break the iteration to keep $\hat p_y(\tilde{x}^s)>c$. We call this additional step \emph{sensible reversion}. Compared with the PGD method, this reversion step requires no additional forward or backward propagation. Therefore, our augmentation is as fast as the PGD method or even faster whenever a reversion is conducted. The proposed training algorithms for both the $\ell_\infty$-norm and the other $\ell_p$-norms are provided in Algorithms 1 and 2. Algorithm 2 is a straightforward extension of Algorithm 1 from the $\ell_\infty$ norm to the $\ell_p$ norm. We note that when the main hyperparameter $c$ is set to 0, the sensible adversarial augmentation is identical to the PGD method so that SENSE-AT generalizes R-AT. Furthermore, existing deep learning implementations that adopt R-AT can be easily extended to adopt SENSE-AT, which requires only inserting an additional sensible reversion step (lines 6-8 in Algorithms 1 and 2). For the practical range of $c$, we recommend to choose $c\in [0.5,1]$ so that any incorrectly classified examples are not allowed to have perturbations. A possible alternative option of the break after the reversion step is to keep the optimization with an added random noise instead of breaking the iteration in order to generate adversarial examples having the loss closer to $-\log c$.

\begin{algorithm}[t]
\caption{Sensible adversarial training for $\ell_\infty$ norm restriction}\label{algo:alg1}
\begin{algorithmic}[1]
\STATE\textbf{Input:} Initialized $f=f_\theta$, $c\in[0.5,1]$, step number $K$, step sizes $\eta_1$, $\eta_2$, data $X_{adv}^{(0)}=X$
    \REPEAT \STATE{\textbf{for} {$i=1,...,m$}, s.t. $f(x_{i,adv}^{(0)})=y_i$}
           \STATE{{~~~~~}\textbf{for} {$k=1,...,K$}}
            \STATE{{~~~~~}{~~~~~}$x_{i,adv}^{(k)}\leftarrow \Pi_{B(x_i,\epsilon)}(\eta_1 sign(\nabla_{x}\ell(f(x_{i,adv}^{(k-1)}),y_i))+x_{i,adv}^{(k-1)})$\text{, $\Pi$: the projection operator}} 
                    \STATE{{~~~~~}{~~~~~}\textbf{if} {$\ell(f,x_{i,adv}^{(k)}, y_i)>\log\frac{1}{c}$ }}
                        \STATE {~~~~~}{~~~~~}{~~~~~}(sensible reversion) $x_{i,adv}^{(K)} = x_{i,adv}^{(k-1)}$ 
                        \STATE {~~~~~}{~~~~~}{~~~~~}\textbf{break}
        \STATE{$\theta\leftarrow\theta -\eta_2 
   \sum_{i=1}^m\nabla_\theta\ell(f,x_{i,adv}^{(K)},y_i)/m$}
    \UNTIL{training converged}\end{algorithmic}
\end{algorithm}

\begin{algorithm}
\caption{Sensible adversarial training for $\ell_p$ norm restriction}
 \begin{algorithmic}[1]
\STATE\textbf{Input:} Classifier $f=f_\theta$, $c\in[0.5,1]$, step number and sizes $K,\eta_1$, $\eta_2$, data $X_{adv}^{(0)}=X$
    \REPEAT \STATE{\textbf{for} {$i=1,...,m$}, s.t. $f(x_{i,adv}^{(0)})=y_i$}
           \STATE{{~~~~~}\textbf{for} {$k=1,...,K$}}
            \STATE{{~~~~~}{~~~~~}{~~~~~}$x_{i,adv}^{(k)}\leftarrow \Pi_{B_p(x_i,\epsilon)}(\eta_1\frac{\nabla_{x}\ell(f(x_{i,adv}^{(k-1)}),y_i)}{\|\nabla_{x}\ell(f(x_{i,adv}^{(k-1)}),y_i)\|_p}+x_{i,adv}^{(k-1)})$\text{, $\Pi$: the projection operator}} 
                    \STATE{{~~~~~}{~~~~~}{~~~~~}\textbf{if} {$\ell(f,x_{i,adv}^{(k)}, y_i)>\log\frac{1}{c}$ }}
                        \STATE {~~~~~}{~~~~~}{~~~~~}{~~~~~}(sensible reversion) $x_{i,adv}^{(K)} = x_{i,adv}^{(k-1)}$ 
                        \STATE {~~~~~}{~~~~~}{~~~~~}{~~~~~}\textbf{break}
        \STATE{$\theta\leftarrow\theta -\eta_2 
   \sum_{i=1}^m\nabla_\theta\ell(f,x_{i,adv}^{(K)},y_i)/m$}
    \UNTIL{training converged}\end{algorithmic}
\end{algorithm}

\subsection{Sensible adversarial training example and loss with implicit loss truncation} Now we discuss the training version of sensible adversarial examples in Definition \ref{def:3} and loss in Definition \ref{def:4}. Through three different rewritings of the sensible adversarial training loss in (\ref{eq:sense_loss2}), we show that (1) the modification effectively renders the sensible adversary framework into practice in the absence of the true Bayes rule, (2) the optimization is conducted through implicit loss truncation that solves zero gradient problem of the modified loss, (3) the optimization is computationally efficient and simple.

\begin{figure}\label{fig:lossAbstrac}
	\centering
	\includegraphics[width=\linewidth]{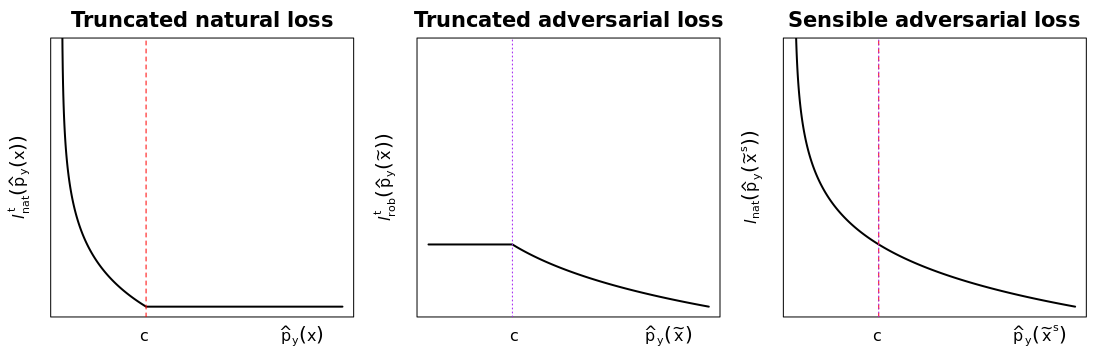}
		\vskip -0.1in
  \caption{The sensible adversarial training loss decomposition. Left: The truncated natural loss $\ell^t_{\rm nat}(\hat p_y(x))$; Middle: The truncated adversarial loss is $\ell^t_{\rm rob}(\hat p_y(\tilde{x}))$;
  Right: The sensible loss is equivalent to the sum of them such as $\ell^{s}(f,x,y)=-\log \hat p_y(\tilde{x}^s)=\ell^t_{\rm nat}(\hat p_y({x}))+\ell^t_{\rm rob}(\hat p_y(\tilde{x}))$. 
  }\label{fig:sen_loss}
  \end{figure}

The first identity will show that the modified loss in (\ref{eq:sense_loss2}) is the sum of a truncated natural loss and robust loss. This explains why this modification can deal with the absence of the true Bayes rule and train a robust classifier. 
Consider a truncated natural loss
$$\ell^t_{\rm nat}(\hat p_y(x))=\max\Big(0, -\log \hat p_y(x)-\log\frac{1}{c}\Big),$$ and a truncated adversarial loss $$\ell^t_{\rm rob}(\hat p_y(\tilde{x}))=\min\Big(-\log \hat p_y(\tilde{x}),\log\frac{1}{c} \Big).$$ Then, we can show that 
\begin{align}
\ell^{s}(f,x,y)&= \ell^t_{\rm nat}(\hat p_y(x))+\ell^t_{\rm rob}(\hat p_y(\tilde{x})).\label{equ:trunc1}
\end{align}
We relegate the derivation to Appendix \ref{app:loss_dev}. The decomposition in (\ref{equ:trunc1}) shows the training sensible loss $\ell^s(f,x,y)$ in Definition \ref{def:4} is equivalent to the sum of  a truncated natural loss and a truncated regular adversarial loss, up to a constant addition for the continuity at the two edges of the truncated area. This is illustrated in Figure \ref{fig:sen_loss}. Although one loss takes $\hat p_y(x)$ as input and the other $\hat p_y(\tilde{x})$, (\ref{equ:trunc1}) is a valid equation because $\hat p_y(x)$ and $\hat p_y(\tilde{x})$ can be calculated from the input of $\ell^s$, that is, $(f,x,y)$. Note that the threshold value 0 in $\ell^t_{\rm nat}$ acts as the minimum of the loss. This $\ell^t_{\rm nat}$ is a surrogate loss of the 0-1 loss to obtain an estimated Bayes rule, as the standard natural loss is, but with a truncated part. On the other hand, the second term can also be seen as a smoother version of the adversarial loss in (\ref{equ:robust_risk}). In this case, the truncated part upper bounds the loss, regularizing influential adversarial examples. Therefore, in SENSE-AT, the truncated adversarial loss prevents adversarial examples from negatively influencing natural accuracy. \emph{In conclusion, the first term forces $f$ to learn the Bayes rule, while the second term to learn a robust model.} 

Finally, we remark that if $c=1$, the sensible loss in (\ref{eq:sense_loss2}) is equivalent to the standard natural loss, and when $c\rightarrow 0$, it becomes the regular adversarial loss. Therefore, the sensible adversarial training loss generalizes the natural loss and adversarial loss. 

 The second identity will show how the truncation happens without explicit truncation of the loss functions. By selectively applying the (untruncated) losses based on a partition of the input space, we can have the same effect as if the truncated losses are applied as in (\ref{equ:trunc1}). In this paper, we use the terminology \emph{implicit loss truncation} to refer to a technique that evaluates a truncated loss by using other \emph{untruncated} losses, which significantly eases the difficulty in optimization. Here we first partition the input space and then see how $\ell^s$ in (\ref{eq:sense_loss2}) applies a different (untruncated) loss on each partition. Let us divide the data into three mutually exclusive groups: 
\begin{align*}
    A_f &= \Big\{ (x,y): \hat p_y(x)\leq c\Big\},\\
    B_f &= \Big\{ (x,y): \hat p_y(x)>c,\hat p_y(\tilde{x})\leq c\Big\},\\
    C_f &= \Big\{ (x,y):\hat p_y(\tilde{x})>c\Big\},
\end{align*}
where $\tilde x$ is the regular adversarial example of $x$. 

Now we apply the $\ell^s$ in (\ref{eq:sense_loss2}) on each partition. First, when $(x,y)\in A_f$, we know that 
\[-\log \hat{p}_y(\tilde{x})\geq -\log \hat{p}_y(x)\geq -\log c.\] Therefore, a data point in $A_f$ has a high natural loss, and its robust loss is truncated. So,
\begin{align*}
\ell^s(f,x,y) &= \ell^t_{\rm nat}(\hat p_y(x)) + \ell^t_{\rm rob}(\hat p_y(\tilde{x}))\\
& =\max\Big(0, -\log \hat p_y(x)-\log\frac{1}{c}\Big) + \min\Big(-\log \hat p_y(\tilde{x}),\log\frac{1}{c} \Big)\\
& = -\log \hat{p}_y(x)=\ell(f(x), y).
\end{align*}
Calculating only the natural loss on $A_f$ is the same as calculating the sum of the two truncated loss on $A_f$. Second, when $(x,y)\in B_f$, we know that
 \[-\log \hat{p}_y(\tilde{x})   \geq -\log c\geq -\log \hat{p}_y(x).\] Therefore, data points in $B_f$ correspond to the points of which both natural and adversarial loss are truncated in (\ref{equ:trunc1}). So,
\begin{align*}
\ell^s(f,x,y) &= \ell^t_{\rm nat}(\hat p_y(x)) + \ell^t_{\rm rob}(\hat p_y(\tilde{x}))\\
& =\max\Big(0, -\log \hat p_y(x)-\log\frac{1}{c}\Big) + \min\Big(-\log \hat p_y(\tilde{x}),\log\frac{1}{c} \Big)\\
&=  \log \frac{1}{c}.
\end{align*}
Putting simply a constant $\log\frac{1}{c}$ on $B_f$ is the same as calculating the sum of the two truncated losses on $B_f$. Likewise, when $(x,y)\in C_f$, we know that \[-\log c\geq-\log \hat{p}_y(\tilde{x}) \geq-\log \hat{p}_y(x).\] Therefore, data points in $C_f$ correspond to the points of which the natural loss is truncated. We have
\begin{align*}
\ell^s(f,x,y) &= \ell^t_{\rm nat}(\hat p_y(x)) + \ell^t_{\rm rob}(\hat p_y(\tilde{x}))\\
& =\max\Big(0, -\log \hat p_y(x)-\log\frac{1}{c}\Big) + \min\Big(-\log \hat p_y(\tilde{x}),\log\frac{1}{c} \Big)\\
&= -\log \hat p_y(\tilde{x})= \ell(f(\tilde{x}), y).
\end{align*}
Calculating only the adversarial loss on $C_f$ is the same as calculating the sum of the two truncated losses on $C_f$. We can write the three cases into the second identity
\begin{align}
\ell^{s}(f,x,y)&= \ell(f(x),y)1_{A_f}+\log \frac{1}{c}1_{B_f}+\ell(f(\tilde{x}),y)1_{C_f}.\label{eq:id_1}
\end{align}
In summary, the sum of the truncated losses in (\ref{equ:trunc1}) can be implemented by applying the natural loss function on $A_f$, a constant loss function $-\log \frac{1}{c}$ on $B_f$, and the adversarial loss function on $C_f$, where none of the functions has explicit truncation on it.

 
 The third identity will emphasize that in SENSE-AT, the implicit truncation in (\ref{eq:id_1}) happens through a single unified input $\tilde{x}^s$, not two inputs as in (\ref{equ:trunc1}) or (\ref{eq:id_1}); the difference in the losses on each partition is leveraged into the adaptiveness of the perturbation ball of $\tilde{x}^s$. Note that the sensible adversarial training example formulation in Definition \ref{def:3} has in itself three types of perturbation balls on different regions of the input space. First, for $(x,y)\in A_f$, $S_{x,\epsilon}(f)=\phi$. Second, for $(x,y)\in B_f$, $S_{x,\epsilon}(f)$ is adaptively decided for each $(x,y)$. Third, the $S_{x,\epsilon}(f)$ for $(x,y)\in C_f$ is the full $\epsilon$-ball centered at $x$. Therefore, with the partition of the space by $A_f,B_f$ and $C_f$, $\tilde{x}^s$ in Definition \ref{def:3} is identical to 

 \begin{equation}\label{eq:sense_def2}
\tilde{x}^s=\left\{
\begin{array}{ll}
x, & \textrm{if } (x,y)\in A_f,\\
\uargmax{z\in S_{x,\epsilon}(f)} \ell (f(z),y), & \textrm{if }(x,y)\in B_f,\\
\tilde{x}, & \textrm{if } (x,y)\in C_f.
\end{array}
\right.
\end{equation}
The sensible adversarial example in (\ref{eq:sense_def2}) has three different stages: a natural $(A_f)$, an adapted $(B_f)$, and a full $\epsilon$-adversarial $(C_f)$ stage. Especially in the adaptive stage $B_f$, the constant loss $\log \frac{1}{c}$ previously in (\ref{eq:id_1}) is implemented by \begin{equation}\label{equ:middle}
\log \frac{1}{c} 1_{B_f}={ \max_{\substack{z\in B(x,\epsilon)\\ \hat p_y(z)\geq c}}}\ell(f(z),y)1_{\{\hat p_y(x)>c,\hat p_y(\tilde{x})\leq c\} },
\end{equation} and $\tilde{x}^s$ is optimized until it has the loss values (approximately) identical to $-\log c$, i.e., $\hat p_y(\tilde{x}^s)= c$. This is the key part of the following third identity:   
 \begin{align}
\ell^{s}(f,x,y)= \ell_{\rm nat}(\hat p_y(\tilde{x}^s)),\label{eq:id_3}
\end{align}
where we use  $\ell_{\rm nat}(\hat p_y)=-\log \hat p_y $ to denote the cross-entropy loss. By this, we see that the three losses in (\ref{eq:id_1}) are unified into a single usual cross-entropy loss by using a single sensible adversarial training example. 
Still, the implicit truncation takes place because of the definition of $ \tilde{x}^s$ in (\ref{eq:sense_def2}), i.e., the adaptive perturbation ball of $\tilde{x}^s$. 

Now we discuss the necessity of the implicit truncation by using the sensible adversarial training example as in  (\ref{eq:id_3}), not with other kinds of examples as in (\ref{equ:trunc1}) or (\ref{eq:id_1}). The decomposition in (\ref{equ:trunc1}) suggests a naive way to optimize (\ref{eq:sense_risk2}) with $x$ and a full PGD example $\tilde{x}$, instead of $\tilde{x}^s$. For each $x$, one can feed the classifier with both $x$ and $\tilde{x}$, and calculate $\ell^s(f,x,y)$ with the explicitly truncated losses in (\ref{equ:trunc1}), and then optimize the empirical risk. However, such an approach is difficult to optimize in practice. Deep learning with a truncated loss is not easy in general, mainly because the truncated area has zero gradients. It is uncertain if the usual gradient descent (GD) method can find a good minimum. Intuitively, when both natural and adversarial losses of a data point are truncated, this point cannot provide any useful information for training. As an extreme case, for a model of which the losses of all points are truncated, there is no way to escape this state by GD. For more details about deep learning with a truncated loss, see \cite{zhang2018generalized}. Meanwhile, if (\ref{eq:id_1}) were directly used, no losses would be explicitly truncated. However, for the points in $B_f$, there is no gradient available of the constant loss function as $-\log\frac{1}{c}$, and the same difficulty arises again. On the contrary, with the implicit truncation in (\ref{eq:id_3}), the loss has the gradient of the cross-entropy loss at any point in this stage. Therefore, this implicit truncation makes the optimization of the empirical risk much easier. 

Next, we discuss what kind of robustness SENSE-AT achieves. Note that SENSE-AT promotes more points to be in $C_f$; The value of $\ell^{s}(f,x,y)$ that is smaller than $-\log c$ is only achieved by any $x\in C_f$. In other words, in $\ell^{s}(f,x,y)$, the points in $A_f$ and $B_f$ are more strongly penalized than those in $C_f$. This is paradoxical because in $C_f$, the strongest full $\epsilon$-ball perturbations are added. Therefore, to obtain a small empirical risk, $f$ should be trained to have all of $x$ to be in $C_f$, and at that moment, $f$ is being trained with the data perturbed on full $\epsilon$ balls, that is, the PGD adversarial examples. Note that if all the data points in SENSE-AT are in the stage $C_f$, then SENSE-AT is identical to R-AT. It is known that the highest robustness has been achieved by training a classifier with R-AT \citep{carlini2019evaluating}. However, if the data and model class do not provide such a favorable condition, i.e., $f$ cannot afford all data points to have full perturbation balls, $f$ is trained to have more data to be in $C_f$ as possible. Note that even for the points in $B_f$, the model is trained to be robust against adversarial perturbations on the adapted $\epsilon$-balls. Although smaller than the full $\epsilon$ ball, it will train $f$ to have the robust area of a truncated point as large as possible because of the adaptive augmentation in (\ref{equ:middle}).

Note that adopting one unified example matters. In SENSE-AT with (\ref{eq:id_3}), the evaluation and  differentiation of the loss need only one forward/backward pass for each $x$, while the naive approach that feeds both $x$ and $\tilde{x}$ needs two times more forward/backward passes. Therefore, by using $\tilde{x}^s$, we save two times of computational complexity. For the same reason, SENSE-AT is faster than other AT methods that deal with the trade-off by using evaluations on both natural and adversarial examples, e.g., \cite{zhang2019theoretically, Wang2020Improving}.

\subsection{Model capacity}
We discuss how SENSE-AT can deal with a limited model capacity. Handling the lack of model capacity is an important matter in the problem of adversarial trade-off. Even when adversarial examples cannot cross the Bayes decision boundary, the trade-off can happen because of the lack of model capacity. Note that given an insufficient model capacity, adversarial training examples of small $\epsilon$ actually can negatively influence the natural accuracy without any true class change. For example, the natural margin of the SOTA models of CIFAR10 significantly decreases by R-AT. Furthermore, due to the lack of model capacity, there could be a subset of data that must be non-robust for the sake of standard accuracy (for example, see the analysis on the Three Clusters dataset in the next section). Adversarial perturbations on these points also can reduce the natural accuracy.

SENSE-AT helps networks with the limited model capacity to obtain their best robustness without collapsing. Note that SENSE-AT inherently involves a set selection mechanism, in which the loss is regularized by the implicit truncation of the adversarial examples for data points in $A_f$ and $B_f$. This truncation helps handle the lack of model capacity because this mechanism prevents any full $\epsilon$-attack from overpowering the learning process. Due to this, our sensible training does not collapse when R-AT collapses, as long as the model capacity is enough for NT as demonstrated in the next section.

\section{Numerical Results}

\begin{figure}
    \centering 
    \includegraphics[
    scale=0.45]{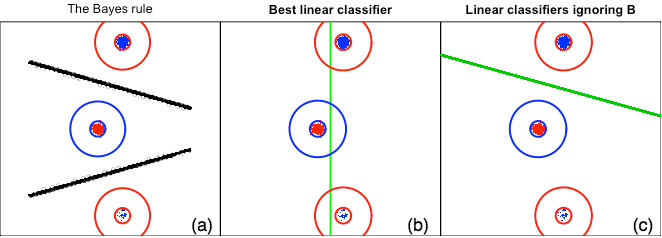}
    \caption{Example 1 with $\sigma=0.2,m=7, \gamma=8$ and $p=0.95$. The inner circle lines denote the 99.9\% quantile region of each cluster, and the outer circle lines are the region that can be reached from the inner circle by $\epsilon$. (a) The Bayes classifier with $R_{\rm adv}(f^B)< 0.001$. (b) Note that if $\gamma=8$, then 99.8\% of data points can cross the decision boundary of the best linear classifier, i.e. $\mathcal{R}_{\rm rob}(f^B)\approx 0.998$. (c) The optimal linear classifier w.r.t. the standard risk when ignoring the bottom cluster.}
    \label{fig:three_normal}
\end{figure}

In this section, we investigate the effectiveness of the proposed method (1) in obtaining high natural and adversarial accuracies and (2) in adversarially training a small model without collapse. Furthermore, we study the sensitivity of our main parameter, the threshold $c$.

We first analyze a synthetic dataset, Three Clusters, to address the trade-off observed in the literature even when a small $\epsilon$ is used, which cannot change the true class of a perturbed example. Then, we move on to real data analysis. We train SENSE-AT models by Algorithm 1 with the $\ell_\infty$-norm on the MNIST \citep{lecun2010mnist} and CIFAR10 datasets \citep{krizhevsky2009learning} with $\epsilon=0.3$ and  $\epsilon=8/255$, respectively. Here the MNIST database is a large database of handwritten digits $0\sim 9$ commonly used to train various image processing systems. The MNIST database contains $70,000$ $28\times 28$ grey images. The MNIST dataset is available at \url{http://yann.lecun.com/exdb/mnist/}. Instead, the CIFAR10 is a more complicated dataset that consists of $60,000$ $32\times 32$ colored images in $10$ classes. The CIFAR10 dataset is available at \url{https://www.cs.toronto.edu/~kriz/cifar.html}. Based on the CIFAR10 dataset, we have performed extensive comparative studies between SENSE-AT and other competitive methods.


Our results confirm that: (1) Sensible training obtains high natural accuracy while keeping comparable adversarial accuracy; (2) Sensible training adversarially  trains a small model without collapsing; (3) Sensible training is not sensitive to its main hyperparameter $c$. These facts suggest that sensible training is a better choice for defense when the model capacity is not enough and we want to balance between natural and adversarial accuracies.

\begin{figure}

	\begin{center}
	\includegraphics[width=.5\linewidth, height=5cm]{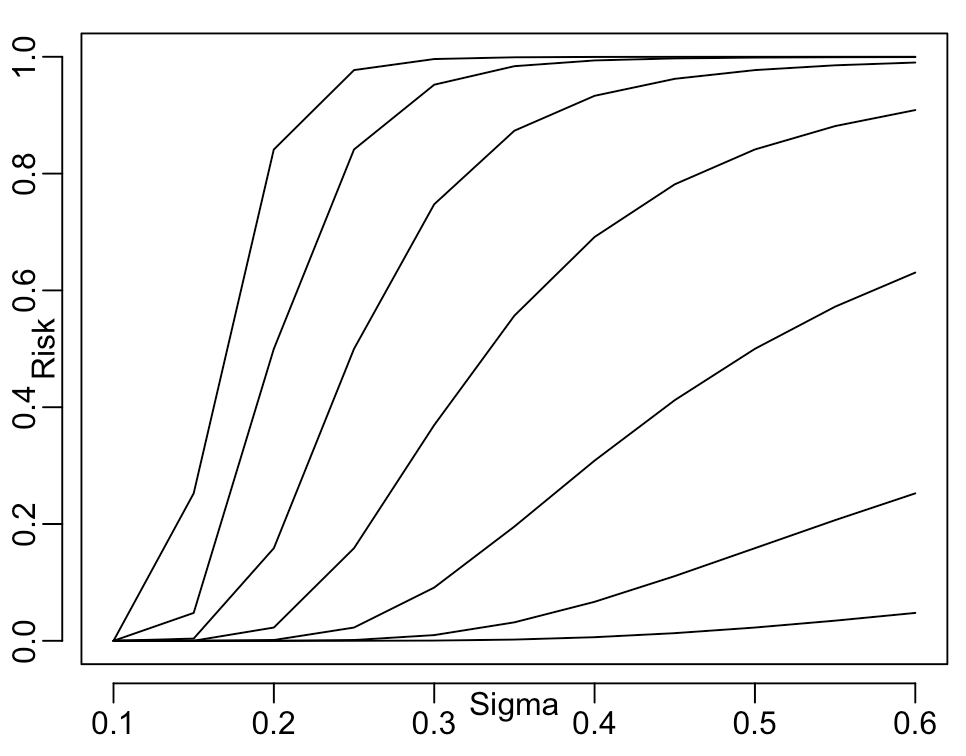}
	\end{center}
	\caption{The adversarial risk in comparison to standard risk of the model $f(x)=sign(x_1)$. The adversarial risk rapidly increases when $\gamma$ increases from 0 to 6 by 1 (bottom to top). Note that $\gamma=0$ corresponds to the standard risk (bottom).}	\label{fig:risk}
\end{figure}

\subsection{Three Clusters dataset} Here we analyze a synthetic dataset that suggests that the SOTA models, which show the trade-off, may not have enough model capacity capable of having both high robust and natural accuracy. This is motivated by Theorem \ref{thm:1}, which sheds light on where the trade-off happens, particularly with small $\epsilon$. If the Bayes decision boundary can be far from data manifolds at least by $\epsilon$, our pursuit of sensible robustness does not cost any adversarial robust risk. That is, adversarial learning is identical to sensible adversarial learning. For this case, if the adversarial perturbations are restricted to $\epsilon$, the robust trade-off may be unnecessary, given enough model capacity such that $f^B\in \mathcal{F}.$ As an example that an insufficient model capacity leads to the trade-off, we introduce the Three Clusters dataset. We first give a general description of the synthetic dataset, then we fix parameters to conduct  the analysis on both adversarial learning and training levels. Then, we will conduct training with both SENSE-AT and R-AT and compare the results.

Consider a uniform binary $y\in\{-1, +1\}$ and three clusters of $X\in\mathbb{R}^2$ such that 
\[
X|y=+1\sim pN(\mu_A,\sigma^2)+(1-p)N(\mu_B,\sigma^2) \mbox{~~~and~~~} X|y=-1\sim N(\mu_C,\sigma^2),
\]
where $\mu_A=(1,m),\mu_B=(1,-m)$ and $\mu_C=(-1,0)$. Assume $p\geq 0.5$. Our attack model is $\ell_2$ restricted adversarial perturbations $\delta$ s.t. $\|\delta\|_2\le \epsilon$. We set $\epsilon=\gamma \sigma$, where $\gamma\geq 2z_{\alpha}+\sqrt{4+m^2}/\sigma$ for $\alpha\in(0,0.5)$ and $z_{\alpha}$ is the upper $\alpha$-quantile of $N(0,1)$. Figure \ref{fig:three_normal}(a) shows the Bayes classifier $f^B$, and the Bayes classifier is robust for a wide range of $\epsilon$ as illustrated in the figure. Details of the derivation are provided in the Appendix \ref{app:three}.

Next, we restrict the model class to linear classifiers denoted by ${\cal F}_{\rm{lin}}$. Note that $f^B\notin {\cal F}_{\rm{lin}}$. The optimal classifier w.r.t. the standard risk is $f_{\rm std}(x)=sign(x_1)$, which is the green line in Figure \ref{fig:three_normal}(b). It is straightforward to show that the standard risk of $f$ is ${\cal R}_{\rm std}(f_{\rm std})=\Phi(-\frac{1}{\sigma^2})$.
Therefore, as $\sigma^2\rightarrow 0$, the standard error ${\cal R}_{\rm{std}}(f_{\rm std})\rightarrow 0$ as well. However, the adversarial risk is
${\cal R}_{\rm{rob}}(f_{\rm{std}}) = \Phi(\gamma-\frac{1}{\sigma^2})$. 
Note that the adversarial risk increases rapidly as $\gamma$ increases (see Figure \ref{fig:risk}). Therefore, $f_{\rm{std}}$ is not a robust model.
 Now, we see that with this model class, the trade-off is inevitable; we must trade natural accuracy for robustness whenever we pursue more robustness than that of $f_{\rm{std}}$, which is the unique minimizer of ${\cal R}_{\rm{std}}(f)$ in $\mathcal{F}_{\rm{lin}}$. For example, let us consider to ignore the bottom cluster in Figure \ref{fig:three_normal}(a) in order to obtain a more robust model in $\mathcal{F}_{\rm{lin}}$. 
%
%
In this case,  the decision boundary is $x_2=-2x_1/m+m/2-\sigma^2\log p/m$. 
We plot these lines in Figure \ref{fig:three_normal}(c) for various $p$ values, which are almost indistinguishable each other. In addition, for a large $\epsilon$, the linear line is robust w.r.t. the two top clusters. Therefore, with the cost of the standard risk $(1-p)/2$ that comes from ignoring the bottom cluster, the linear model improves the robustness that is upper bounded by $p/2$.

We fix the parameters as $\sigma=0.2,m=7, \gamma=8$ and $p\in[0.5,1)$. First, we consider the theoretical side with 0-1 loss. In fact, in this case, more than 99.9\% of data points do not cross the Bayes decision boundary. Therefore, almost all of the sensible adversarial examples are identical to the regular adversarial examples, and thus $|R_{\rm rob}^{s,B}(f)-R_{\rm rob}(f)|<0.001$. Therefore, the sensibly optimal models will behave very similar to the regular adversarially robust models, losing $1-p$ standard accuracy. This seems to contradict to our theory that shows the sensibly optimal model is the Bayes rule. This gap comes from that the theory assumes an enough model capacity that can have the Bayes rule in it, but the case we are discussing does not have such an enough model capacity. Now we consider the algorithmic side with the cross entropy loss. For R-AT, it is difficult to tell if it will collapse to a constant function or not. In its algorithm, R-AT uses the cross entropy loss, which is unbounded on miss-classified adversarial data points. Note that it is impossible to get every adversarial data points to be correctly classified by any linear model in $\mathcal{F}_{\rm{lin}}$. Therefore, it is hard to say if the R-AT model would be similar to Figure \ref{fig:three_normal} (b) or Figure \ref{fig:three_normal} (c). In SENSE-AT, models like Figure \ref{fig:three_normal} (c) has highly dominating natural loss for the cluster $B$. However, models like Figure \ref{fig:three_normal} (b) will have a loss bounded by $-\log c.$ Therefore, we anticipate that the optimal SENSE-AT models will be similar to the model in Figure \ref{fig:three_normal} (b). In conclusion, when the model capacity is insufficient to achieve robustness, sensible adversarial training will prioritize keeping the natural performance. 

\begin{figure}

	\begin{center}
	\includegraphics[width=.35\linewidth]{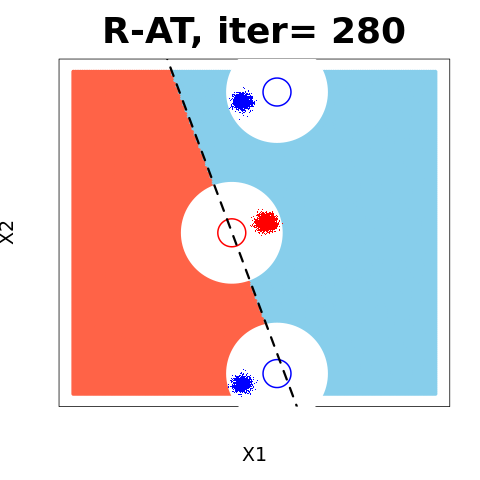}
	\includegraphics[width=.35\linewidth]{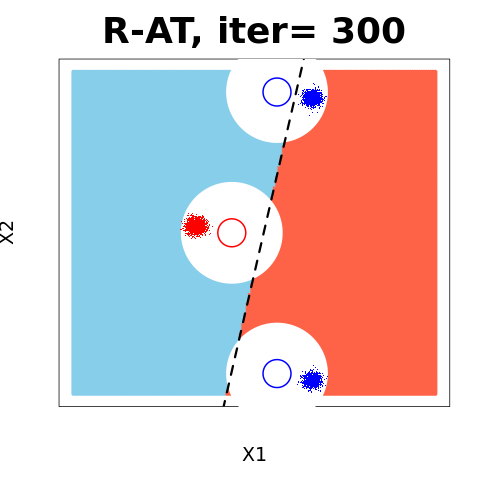}
	\includegraphics[width=.35\linewidth]{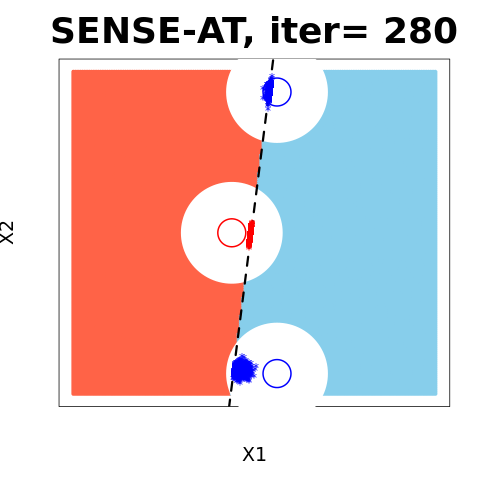}
	\includegraphics[width=.35\linewidth]{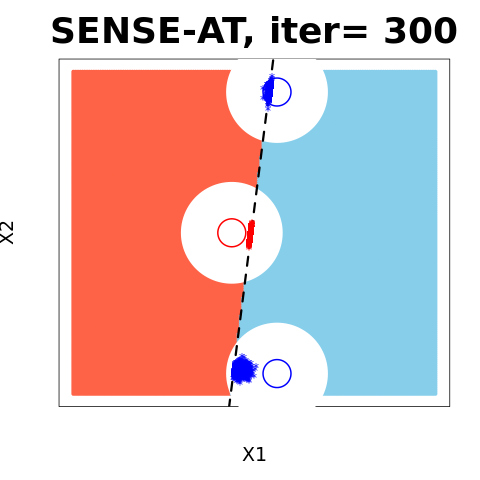}
	\end{center}
	\caption{Top: the model's decision boundary by R-AT alternates between two different types. The blue and red points are the augmented adversarial training examples w.r.t the current model. Bottom: in SENSE-AT, the decision boundary converges to a model similar to the optimal linear classifier. The blue and red points are the augmented sensible adversarial training examples w.r.t the current model. }\label{fig:toy_emp}
\end{figure}
For $p=0.55,$ we actually conduct SENSE-AT and R-AT on the simulated data and confirm the previous  analysis on the ATs. As the simulation requires a light computation, we implement it with the R programming language. Because the model is essentially linear, the PGD is set $K=1$ with $\epsilon = \gamma \times \sigma= 1.6$. The hyperparameter for SENSE-AT is set $c=0.9$. Both R-AT and SENSE-AT are initialized with a linear model fitted by logistic regression. We train with a learning rate of $0.01$ for 300 iterations. The results are illustrated in Figure \ref{fig:toy_emp}. The training with R-AT did not converge, but the model's decision boundary keeps alternating between the two types shown in Figure \ref{fig:toy_emp}. On the other hand, SENSE-AT stably converges from the early iterations and keeps the decision boundary similar to that of the optimal linear classifier as in Figure \ref{fig:three_normal} (b). We see that the trained model in SENSE-AT shows some trade-off; although the natural accuracy is not 100\%, it prevents the natural accuracies are completely neglected while obtaining some robustness when the model capacity is not enough.

{ As a final note, we mention that this example also shows a collapse scenario as an extreme case.} Consider when $\epsilon$ is large enough to cross the decision boundary. Let $\alpha\rightarrow 0$. Then, in order to obtain any $\varrho$ robust accuracy with any $\varrho>0$, the entire natural accuracy cannot exceed $0.5+\varrho$. Therefore, regardless of the given model capacity, one of the approximately most robust classifiers is a constant function either $f(x)=+1$ or $f(x)=-1$ for all $x$, achieving 50\% of adversarial accuracy. { In the following experiments, we will show how our SENSE-AT prevents the trained model from collapsing in real-data settings.}


\subsection{MNIST dataset} $~~$

\subsubsection{Natural accuracy and adversarial accuracy}
We conduct SENSE-AT on the MNIST dataset \citep{lecun2010mnist}. For the model architecture, we use a convolutional neural network (CNN) with three convolutional layers followed by a fully connected linear layer, which is coded and used by \citep{zhang2019theoretically}. The model is pre-trained with NT and then trained with SENSE-AT with $\epsilon=0.3$, $\eta_1=0.05$ and $K=10$ for $c=0.7$. The initial SGD learning rate is $0.01$, and we train for 500 epochs. The SENSE-AT model is tested with $\ell_\infty$ PGD attacks of $\epsilon=0.3$ with the step number $K=500$ and step size $\eta_1=0.01$ with 20 random restarts by a Python package Advertorch \citep{ding2019advertorch}. The step number $K=500$ is enough for PGD to optimize strong PGD attacks (see, Figure \ref{fig:mnist_PGD_check}). Our model achieves $91.74\%$ accuracy against the PGD attacks and $99.51\%$ natural accuracy.


Now we check the sensitivity to $c$ together with varying model capacity. To this aim, we consider a sequence of CNNs with an increasing number of convolutional kernels. In terms of network capacity, we call a network having {\it capacity $d$} when the network has two convolutional layers with $2^{(d-1)}$ and $2^d$ kernels respectively, followed by a fully connected linear layer of $2^{(d+4)}$ units. Each layer is activated by ReLU $\max(x, 0)$. Each convolutional layer is followed by a $2\times 2$ max-pooling layer. The size of all convolutional kernels is $5\times 5$. We pre-train the CNNs with NT and then train with SENSE-AT with $\epsilon=0.3$, $\eta_1=0.05$ and $K=10$ for varying $c\in \{0.0,0.1,\cdots,0.9\}$. The initial learning rate $\eta_2$ is $0.01$, and we train for 500 epochs. When training the R-AT models, we use the same hyperparameters except $c$.

\begin{figure}
	\centering
	\includegraphics[width=.95\linewidth]{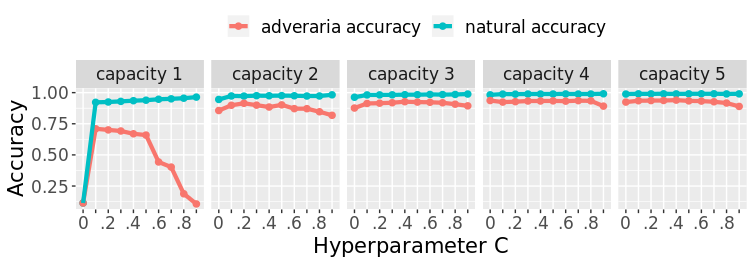}
  	\caption{The natural and robust accuracies of our models for the varying parameter $c$ for MNIST dataset.}
	\label{fig:exp1acc}
	\end{figure}

 By natural training, the networks of capacity 1 and capacity 2 achieve about 95\% and 97\% accuracy, whereas the networks of the other capacities achieve more than 99\%. When trained with R-AT, the networks with capacities 1,2, and 3 collapse \citep{Madry2017Towards}. Therefore, capacity 3 is enough only for NT, and capacities 1 and 2 are possibly insufficient even for NT. The results are in Figure \ref{fig:exp1acc}. Our SENSE-AT models with any capacity did not collapse, except capacity 1 with $c=0.0$, which is identical to R-AT without any perturbation on the naturally incorrectly classified examples. 
 

  
\subsubsection{Convergence check for PGD adversarial attacks}
\begin{figure}[h]
	\begin{center}
	\includegraphics[width=.8\linewidth, height=7cm]{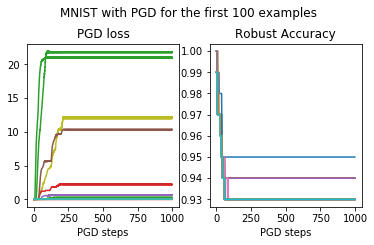}
	\end{center}
	\caption{The convergence check for the PGD attacks on the MNIST model for the step size 0.01. }\label{fig:mnist_PGD_check}
\end{figure}

It is important to check proper optimization in generating the PGD attacks. This has become an almost standard practice to check the convergence of the PGD loss for multiple initializations. We also check the quality of PGD adversarial attacks used to test our SENSE-AT models. The models are tested with $\ell_\infty$ PGD attacks of $\epsilon=0.3$ with the step number $K=500$ and step size $\eta_1=0.01$ by a Python package called Advertorch \citep{ding2019advertorch}. For the quality check, we consider the first 100 test examples. The left panel of Figure \ref{fig:mnist_PGD_check} shows the cross-entropy loss on the perturbed examples by the PGD steps. Each line stands for each original test input, and the total 100 lines are not well distinguishable as the lines overlap. The right panel of Figure \ref{fig:mnist_PGD_check} shows the cumulative worst-case robust accuracy. As the step number increases, the lines are not well distinguishable. In conclusion, we can see that $K=500$ is enough to achieve the lowest point by counting the worst case.

\subsection{CIFAR10 dataset}
Throughout our experiment on CIFAR10, we adopt the warmup and learning schedule and parameters for SGD that are used by \cite{balaji2019instance}. For the warmup period (the first several epochs), we generate full PGD adversarial examples for the correctly classified inputs and have a loss less than the threshold. 
  
\subsubsection{Natural accuracy and adversarial accuracy}\label{exp:cifar1}

We train the SENSE-AT models with ResNet \citep{he2016deep}. ResNet is a widely used DNN architecture designed to alleviate a degradation problem in training deep models. Its shortcut connection between its building blocks explicitly forces the building blocks to learn so-called ``residual mappings". Among many possible ResNet design choices, we train with ResNet-34-10 (a 34-layer ResNet for 10-class classification) \citep{he2016deep}, the architecture used for current SOTA models in literature such as the MART model \citep{Wang2020Improving} and the TRADES model \citep{zhang2019theoretically}. The ResNet is trained by Algorithm 1 of the proposed SENSE-AT, where our main hyperparameter $c$ as set to 0.5 along with other hyperparameters detailed in Table \ref{tb:hyper_cifar}.

\begin{table}[h]
  \caption{The learning specifications for the SENSE-AT models on CIFAR10}
  \label{tb:hyper_cifar}
  \centering
  \begin{tabular}{c|c|c|c}
    \toprule
Hyperparameters&WideResNet-34-10 & ResNet-20,32,44,56 &CNN\\\midrule\midrule
SGD initial learning rate & 0.1 &0.1&0.01\\
Learning rate annealing &Step decay &Step decay&Step decay\\
Learning rate decay steps&[70,90,100]&[80,140,170]&[80,140,170]\\
Learning rate decay factor &0.2&0.1&0.1\\
Weight decay &0.0005 &0.0002& 0.0002\\
Training number of epochs & 300&200&150\\\midrule
Warmup period &10 epochs&5 epochs&5 epochs\\\midrule
$K$&10&10&10\\
$\epsilon$&8/255&8/255&8/255\\
 $a$&2/255&2/255&2/255\\\bottomrule
  \end{tabular}
\end{table}

We test the trained model with adversarial perturbations with $\ell_\infty$-norm less than $\epsilon=8/255$. We generate the attacks with gradient-based attack methods or white-box attacks, such as PGD-K$_{(r)}$ (the PGD attack with K steps with $r$ random restarts), FGSM \citep{FGSM}, and DeepFool \citep{moosavi2016deepfool}. We use the attacks implemented in Foolbox \citep{foolbox}, Advertorch \citep{ding2019advertorch}, and Adversarial Robustness 360 Toolbox (ART) \citep{art2018}. The attack specifications are in Table \ref{attck_sp} and the options that are not listed in the table are kept as default. 

 \begin{table}[ht]
  \caption{The white-box attack specifications. We denote the step size and step number by $\eta_1$ and $K$.}
  \label{attck_sp}
  \centering
  \begin{tabular}{c|c|cc|cl}
    \toprule
Dataset & Attack &$\eta_1$&$K$&Python package &Function\\\midrule
	CIFAR10& PGD100& $\frac{2}{255}$ & 100 &Advertorch&LinfPGDAttack\\
	CIFAR10& PGD40& $\frac{2}{255}$ & 40 &Advertorch&LinfPGDAttack\\
	CIFAR10& DeepFool& default & default &Foolbox&DeepFoolLinfinityAttack\\
	CIFAR10& FGSM& $\frac{8}{255}$&1 &Advertorch&GradientSignAttack\\
	\bottomrule
  \end{tabular}
\end{table}

We also test our models with transfer attacks, which is a type of black-box attack, as robustness against such attacks is important from a security perspective. It is well known that adversarial attacks transfer \citep{good_explaining}, which means that a classifier, so-called a victim classifier, can be fooled by an adversarial attack optimized by using another classifier. This can happen even when the two models do not share the same model architecture, trained by different algorithms. Therefore, a classifier being attacked by a transfer attack is one of the most feasible scenarios in real applications because even without knowing the parameters or structure of the victim model, an attack is still possible. Given a test image, each transfer attack is identified by its generating model, on which the PGD method is applied to optimize the adversarial perturbation. Then we use this attack to other victim models, or defense methods, to test their defense performance.

\begin{table}[t]
  \caption{The test result on natural examples and $\ell_\infty$-attacks for CIFAR10 $(\epsilon=8/255)$. 
  }
  \label{cifar_table}
  \centering
  \begin{tabular}{cccccc}
    \toprule
    &\multicolumn{4}{c}{Test data}\\ \cline{2-5}
   Defense Method &Natural&FGSM&DeepFool&PGD100$_{(5)}$\\
    \midrule
MART &  83.62 &  61.61& 83.34& \textbf{55.81} \\
TRADES & 84.92 & 60.87& 84.68 & 55.40 \\
MMA-12 &88.59 & 60.37& 89.60 & 43.90\\
IAAT & 89.55 & 62.20& 89.25 &50.04\\
SENSE-AT &\textbf{\NaturalTop} & \textbf{65.52}& \textbf{90.60} &\RobustTop 
\\\bottomrule
  \end{tabular}
\end{table}

As baselines, we compare the SENSE-AT model to the TRADES \citep{zhang2019theoretically}, MART \citep{Wang2020Improving}, MMA \citep{Ding2020MMA}, and IAAT \citep{balaji2019instance}. Since \cite{balaji2019instance} did not release their trained model, we reproduced the IAAT model from the code provided by \cite{balaji2019instance}. Any accuracy numbers for the baseline models are calculated by applying the same attacks we generate to test the SENSE-AT models. The results are shown in Table \ref{cifar_table}. We can see that SENSE-AT can achieve the highest natural accuracy while keeping comparable robustness. In Table \ref{cifar_table2}, we test with transfer attacks generated by $\ell_\infty$ PGD40 attacks with $\epsilon=8/255$. The meaning of transfer attacks is that we generate adversarial examples from one model, such as MART, and use it to attack another model, such as SENSE-AT.
The subscript of the column names denotes the corresponding generating model. Overall, the robustness of the SENSE-AT model is comparable to defend against transfer attacks. Given this comparable performance of SENSE-AT, the easy hyperparameter tuning of our method can favor real applications. As we will see in the next section, our method is not sensitive to the choice of the main hyperparameter $c$, which crucially promotes the practicality of our method.

\begin{table}[h]
  \caption{PGD40 transfer attacks on CIFAR10. Ten attack sets are generated by each model. 
  }
  \label{cifar_table2}
  \centering
  \begin{tabular}{lcccccc}
    \toprule
    &\multicolumn{5}{c}{Test data}\\ \cline{2-6}
   Defense method &	PGD$_{\rm MART}$&PGD$_{\rm TRADES}$&PGD$_{\rm MMA-12}$&PGD$_{\rm IAAT}$&PGD$_{\rm SENSE}$\\\hline
    MART& 55.73	&63.43	&70.22&	67.25&	72.84\\
    TRADES & 66.05&	54.82&	{70.82}&	{67.77}&	{73.06}\\
   MMA-12 &69.30&	67.13&	44.04	&64.60&	71.32\\
    IAAT &71.02&	68.32&	70.60&	50.07&	70.86\\
    SENSE-AT & {71.46}&	{68.71}&	70.14&	63.80&	49.51\\
	\bottomrule
  \end{tabular}
  \vspace{-0.2cm}
\end{table}

\subsubsection{Sensitivity analysis on $c$}

With the model architectures used in \ref{exp:cifar1}, we change the main hyperparameter $c$ to investigate the sensitivity. Here we consider the SENSE-AT models trained on CIFAR10 with $c\in\{0.1,0.3,0.5,0.7\}$. Note that the recommended range of $c$ is $c\in[0.5,1]$ to ensure no perturbations on naturally incorrectly classified examples. However, to more thoroughly investigate the sensitivity, we also consider $c<0.5$. We reduce the number of epochs to 200, 200, and 254 for when $c$ is 0.1, 0.3, and 0.7, respectively.

In Table \ref{cifar_table3}, we can observe that $c$ controls the trade-off between natural and robust accuracies. The natural accuracy is positively correlated to the $c$ value, and the minimum robustness accuracy numbers show a negative correlation tendency to $c$. 

\begin{table}[t]
  \centering \caption{The sensitivity of $c$ of SENSE. Tested on the first 1000 test data points.}\label{cifar_table3}
  \begin{tabular}{llllll}
    \toprule
    &&\multicolumn{4}{c}{Hyperparameter c}\\ \cline{3-6}
 Defense method&\text{Test data}& c=0.1 & c=0.3&c=0.5& c=0.7\\
    \midrule
		\multirow{2}{*}{SENSE-AT}&\text{NAT} &89.10&90.30&91.40&91.25\\ 
&\text{PGD}100$_{5}$ &44.70&45.20&50.90& 49.40\\ 
	\bottomrule
  \end{tabular}
  \end{table}

\subsubsection{Sensitivity analysis on model capacity}

We train a sequence of models with increasing model capacity with a fixed $c=0.7$. We train ResNet-20, 32, 44, and 56 \citep{he2016deep}, each of which respectively has 179, 104, 73, and 57 times fewer parameters than WideResNet-34-10. These models are implemented in python code by \cite{Idelbayev18a}. We report the results in Table \ref{cifar_table4}. These models can not be enough for R-AT on CIFAR10. Nonetheless, all the models achieve more than half of the robustness of the SOTA models TRADES and MART in Table \ref{cifar_table}, while maintaining even much better natural accuracy.

\begin{table}[h]
  \centering
  \caption{The SENSE-AT results with small models. Tested on the first 1000 test data points.}\label{cifar_table4} 
  \begin{tabular}{llllll}
    \toprule
    &&\multicolumn{4}{c}{Network}\\ \cline{3-6}
    Defense method&
    \text{Test data}& Res20 &Res32 &Res44& Res56\\
    \midrule
		\multirow{2}{*}{SENSE-AT}&NAT &87.50&88.30&89.30&89.50\\
&PGD100$_{5}$ &26.50&29.50&32.50&34.30\\ 
\bottomrule
  \end{tabular}
\end{table}

   For CIFAR10, we have observed a positive correlation of natural accuracy and a negative correlation of robustness to the threshold hyperparameter $c$. For MNIST, on the other hand, we have observed that as the model capacity increases, the accuracy numbers become almost insensitive to $c$. This result of MNIST suggests that the correlations in CIFAR10 may imply that the model capacity used for CIFAR10 is much smaller than we need to achieve a robust model. Furthermore, if we can assume that $\epsilon=8/255$ is small enough so that any PGD attack can neither escape the original data manifold nor change the original class, then sensible training is equivalent to training with on-manifold data of correct labels. For the truncated loss with data augmentation, a larger $c$ leads to more confident fitting on the observed natural data. Therefore, the decrease in robust accuracy by increasing $c$ indicates that to fit these natural data more confidently, the model needs to give up more unobserved on-manifold points. In other words, for the current model capacity, it may not be possible to achieve a perfect classifier in terms of both natural and adversarial accuracy. 
  
 

\subsubsection{Model collapse}

Here we try to train a robust model with a small capacity. We train CNNs with $c\in\{0.1, 0.5,0.7\}$ for the training. We use CNN models with two convolutional layers followed by three fully connected linear layers. We consider two capacities: capacity 1 is a network given in the pytorch tutorial for CIFAR10 \footnote{\url{https://pytorch.org/tutorials/beginner/blitz/cifar10_tutorial.html}}, and capacity 1 has 4 times fewer channels than capacity 2 for the linear layers. For NT and R-AT, we use the same learning hyperparameters except $c$. When capacity 1 is used, we trained with various SGD learning rates 0.01, 0.05 and 0.001 for R-AT to enable adversarial training. However, for all of those learning rates, R-AT collapsed. We test the trained models with PGD100 with random 5 restarts. The test result is presented in Table \ref{cifar_table_2}. Given the serious lack of model capacity, the decrease in natural accuracy of SENSE-AT models compared with the naturally trained model is not serious while achieving better robust accuracy numbers. Note that none of our models collapse, and therefore the learnability of sensible adversarial training is not sensitive to the choice of $c$ even in the lack of model capacity.

\begin{table}
 \caption{The test result on the CNNs of capacity 1 and capacity 2. Tested on the first 1000 test data points.}\label{cifar_table_2}
  \centering
  \begin{tabular}{cclllll}
    \toprule
Network &Test data& R-AT &c=0.1 & c=0.5& c=0.7 &NT\\
    \midrule
\multirow{2}{*}{Capacity 1}&		NAT &collapse&57.9&64.1&67.4&68.8\\ 
&PGD&collapse &25.9&6.9&0.7&0.0\\
    \midrule
\multirow{2}{*}{Capacity 2}&		\text{NAT} &49.1&61.5&70.19&72.1&76.0\\ 
&\text{PGD} &31.5&27.6&11.2& 4.7&0.0\\ 
	\bottomrule
  \end{tabular}
 \end{table}
 
 \subsubsection{Convergence check for PGD adversarial attacks}
\begin{figure}
	\begin{center}
	\includegraphics[width=.95\linewidth, height=5cm]{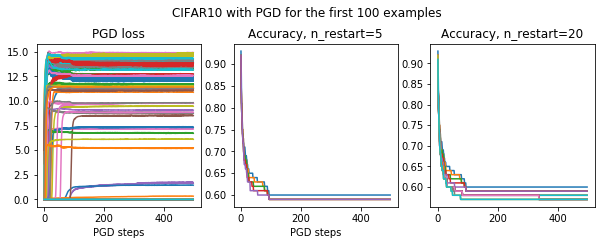}
	\end{center}
	\caption{The convergence check for the PGD attacks on the CIFAR10 model for the step size 2/255. }\label{fig:cifar_PGD_check}
\end{figure}

  For the PGD attacks, we draw accuracy and loss plots for increasing step numbers in Figure \ref{fig:cifar_PGD_check}. We consider the first 100 test examples, and the SENSE-AT model on CIFAR 10 is used (ResNet-34-10). We can see that 100 steps are enough for the PGD optimization. The left panel of Figure \ref{fig:cifar_PGD_check} shows the cross-entropy loss on the perturbed examples by the PGD steps for one random restart. Each line stands for each original test input. The middle panel of Figure \ref{fig:cifar_PGD_check} shows the cumulative worst-case robust accuracy for five random restarts. The line of the highest accuracy numbers stands for the first random restart. The next line stands for the worst-case accuracy between the first and second random restart. Likewise, the line of the lowest accuracy numbers denotes the worst-case accuracy among five random restarts. The right panel of Figure \ref{fig:cifar_PGD_check} shows the cumulative worst-case robust accuracy for 20 random restarts. Compared with the five-random-restart case, the worst performance decreases about $2\%$. We use the step size 2/255 rather than $\epsilon\times 2/ 255$ because it is more effective. 

\section{Concluding remarks}
In this paper, we have proposed a novel sensible adversary which is useful for learning a defense model and keeping high natural accuracy simultaneously. We have theoretically established that the Bayes classifier is most robust under the framework of the sensible adversary. Our training algorithm is efficient and stable and not sensitive to the choice of the main hyperparameter $c$. Furthermore, $c$ has a clear interpretation as the lowest prediction-probability bound.

We have implemented our method on two large-scale image classification datasets: MNIST and CIFAR10. We have performed an extensive comparative study and compared SENSE-AT with other adversarial training methods.
Our empirical experiments yield very competitive results of adversarial training on both MNIST and CIFAR10. SENSE-AT achieves both high natural accuracy and robust adversarial accuracy. 
In addition, we showed that the sensible approach can effectively deal with the lack of model capacity. This is because by paying a robust accuracy on a certain area, the algorithm protects the model from being collapsed by influential adversarial examples.

We now mention several future directions for research on the sensible adversary. One remaining theoretical problem is to develop generalization error bounds for sensible adversarial training to theoretically justify our empirical performance. In this paper, we did not tackle the lack of sample size. Therefore, there is much remaining work to be done to theoretically understand the models trained with sensible adversarial examples related to generalization.

\newpage
\appendix


\section{Proofs of Theorems}

\begin{proof}[{\rm\bf Proof of Theorem 1}]
(a). Observe that
\begin{align*}
    \mathcal{R}_{\rm rob}^s{(f^B)}&=\mathbb{P}(f^B(\tilde{X}^s)\neq Y) \\&=
    \mathbb{P}(f^B(X)\neq Y)+ \mathbb{P}(f^B(X) = Y, \text{and }~\exists x'\in S_{X,\epsilon}(f^B)~ s.t.~ f^B(x')\neq Y)\\ &= \mathbb{P}(f^B(X) \neq Y)+0, ~~~~~~\text{by the definition of } S_{X,\epsilon}(f^B) \\&= \mathcal{R}_{\rm std}{(f^B)}=\mathcal{R}_{\rm std}^*.
\end{align*}

It is obvious that $f^B$ is a minimizer of $\mathcal{R}_{\rm rob}^s(f)$ because $ \mathcal{R}_{\rm std}^*$ is a lower bound of $\mathcal{R}_{\rm rob}^s(f)$ for any $f\in \mathcal{F}$.
Now we show $f^B$ is the unique minimizer of $\mathcal{R}_{\rm rob}^s(f)$. We observe
\begin{align}
    \mathcal{R}_{\rm rob}^s(f)&= {\mathbb{P}}(f^B(X)\neq Y, f(X)\neq Y)+{\mathbb{P}}(f^B(X) = Y, \exists x' \in S_{X,\epsilon}(f^B)~s.t.~f(x')\neq Y)\nonumber
    \\&=\sum_{k=1}^K {\mathbb{P}}(f^B(X)\neq k, f(X)\neq k,Y=k)+{\mathbb{P}}(f^B(X) = k, \exists x' \in S_{X,\epsilon}(f^B)~s.t.~f(x')\neq k,Y=k)\nonumber\\&=\sum_{k=1}^K\int_{{\mathcal{X}}} {\mathbb{P}}(f^B(x)\neq k, f(x)\neq k,Y=k|X=x)\nonumber\\&+{\mathbb{P}}(f^B(x) = k, \exists x' \in S_{x,\epsilon}(f^B)~s.t.~f(x')\neq k,Y=k|X=x)d\mathbb{P}(x)\nonumber\\&=\sum_{k=1}^K\int_{{\mathcal{X}}} 1_{f^B(x)\neq k, f(x)\neq k}\eta(k|x)+1_{f^B(x) = k, \exists x' \in S_{x,\epsilon}(f^B)~s.t.~f(x')\neq k}\eta(k|x)d\mathbb{P}(x)\nonumber\\&=\mathcal{R}_{\rm std}(f)+\sum_{k=1}^K\int_{{\mathcal{X}}} 1_{f^B(x)= k}(1_{ \exists x' \in S_{x,\epsilon}(f^B)~s.t.~f(x')\neq k}-1_{f(x)\neq k})\eta(k|x)d\mathbb{P}(x). \label{eq:Rrobs}
\end{align}
The last equality is by $1_{f^B(x)\neq k, f(x)\neq k}=(1-1_{f^B(x)= k})1_{f(x)\neq k}.$ The first term $\mathcal{R}_{\rm std}(f)$ is uniquely minimized by the Bayes rule $f^B$. The second term is always non-negative, and is zero when $f=f^B$. Therefore, $\mathcal{R}_{\rm rob}^s(f)$ is uniquely minimized by $f^B$.


(b). For any set $A\subset \mathcal{X}\setminus DB(f^B,\epsilon)$, $\mathbb{P}(\exists x' \in B(X,\epsilon)~ s.t.~ f^B(x')\neq Y, X\in A) =\mathbb{P}(f^B(X)\neq Y, X\in A)$. Note that on any subset $B\subset \mathcal{X}$, the Bayes rule has the least error probability. Therefore, $\mathbb{P}(f^B(X)\neq Y, X\in A)\leq \mathbb{P}(f(X)\neq Y, X\in A)\leq \mathbb{P}(\exists x' \in B(X,\epsilon)~s.t.~f(x')\neq Y, X\in A)$. The last inequality is trivial because if $f(X)\neq Y \Rightarrow \exists x' \in B(X,\epsilon)~s.t.~f(x')\neq Y$.
\end{proof}

\begin{proof}[{\rm\bf Proof of Theorem 2}]

The sensible risk of $f$ w.r.t. the restricted distribution corresponding to (\ref{eq:Rrobs}) is \begin{align}
    \tilde{\mathcal{R}}_{\rm rob}^s(f)&=\tilde{\mathcal{R}}_{\rm std}(f)+\sum_{k=1}^K\int_{\tilde{\mathcal{X}}} 1_{f^B(x)= k}(1_{ \exists x' \in S_{x,\epsilon}(f^B)~s.t.~f(x')\neq k}-1_{f(x)\neq k})\eta(k|x)d\tilde{\mathbb{P}}(x)\label{eq:all_part}\\ &=\tilde{\mathcal{R}}_{\rm std}(f)+ \int_{\tilde{\mathcal{X}}}\sum_{k=1}^K1_{f^B(x)= k}(1_{ \exists x' \in S_{x,\epsilon}(f^B)~s.t.~f(x')\neq f^B(x)}-1_{f(x)\neq f^B(x)})\eta(f^B(x)|x)d\tilde{\mathbb{P}}(x)\nonumber\\&=\tilde{\mathcal{R}}_{\rm std}(f)+\int_{\tilde{\mathcal{X}}}(1_{ \exists x' \in S_{x,\epsilon}(f^B)~s.t.~f(x')\neq f^B(x)}-1_{f(x)\neq f^B(x)})\eta(f^B(x)|x)d\tilde{\mathbb{P}}(x),\label{eq:second_part}
\end{align}
because $\tilde{\mathbb{P}}(Y=k|X=x)={\mathbb{P}}(Y=k|X=x)$ for $x\in{\tilde{\mathcal{X}}}$. The minimum of  $\tilde{\mathcal{R}}_{\rm rob}^s(f)$ is achieved by $f^B$ with the minimum value $\tilde{\mathcal{R}}^{*}=\tilde{\mathcal{R}}_{\rm std}(f^B)+0$. Therefore, any function $f$ that $\tilde{\mathcal{R}}_{\rm std}(f) > \tilde{\mathcal{R}}^{*}$ cannot achieve the minimum of $\tilde{\mathcal{R}}_{\rm rob}^s(f)$ because the term $(1_{ \exists x' \in S_{x,\epsilon}(f^B)~s.t.~f(x')\neq f^B(x)}-1_{f(x)\neq f^B(x)})$ in (\ref{eq:second_part}) is always non-negative. Therefore, only functions in $\mathcal{A}=\big\{f\in\mathcal{F}|\tilde{\mathbb{P}}\big(f(X)=f^B(X)\big)=1\big\}$ need to be considered as possible minimizers of $\tilde{\mathcal{R}}_{\rm rob}^s(f)$.

 Note that $\mathcal{A}_\epsilon\subset \mathcal{A}$. By the definition, we know that $f\in\mathcal{A}\setminus\mathcal{A}_\epsilon $
if and only if \begin{align*}
    & i).~~~\tilde{\mathbb{P}}\big(f(X)=f^B(X)\big)=1\\& ii).~~~\tilde{\mathbb{P}}\big(f(x)=f^B(x),~\forall x\in S_{X,\epsilon}(f^B)\big)<1.
\end{align*}
Therefore, for $f\in\mathcal{A}\setminus\mathcal{A}_\epsilon $, $\exists A\subset \tilde{\mathcal{X}}$ s.t. $\tilde{\mathbb{P}}\big(X\in A\big)>0$ and $\exists x'\in S_{x,\epsilon}(f^B)$ for $\forall x\in A$
 s.t. $f(x')\neq f^B(x').$ For this $f$, the equation in (\ref{eq:second_part}) can be written as  $\tilde{\mathcal{R}}_{\rm rob}^s(f)= \tilde{\mathcal{R}}^{*} + \alpha$ for some $\alpha\geq 0$. Now we show that $\alpha>0$.

Note that by the definition of the Bayes rule, $\mathbb{P}(y=f^B(x)|X=x)\geq \frac{1}{K}$. Otherwise, $1=\sum_{k=1}^K\mathbb{P}(y=k|X=x)\leq K\mathbb{P}(y=f^B(x)|X=x)<1$, which is contradict. Then, for the $f\in \mathcal{A}\setminus\mathcal{A}_\epsilon$ and $A\subset\tilde{\mathcal{X}}$ that are described above, 
\begin{align*}
   &\int(1_{ \exists x' \in S_{x,\epsilon}(f^B)~s.t.~f(x')\neq f^B(x)}-1_{f(x)\neq f^B(x)})\eta(f^B(x)|x)d\tilde{\mathbb{P}}(x)\\
   &\geq \frac{1}{K}\int_{\tilde{\mathcal{X}}}(1_{ \exists x' \in S_{x,\epsilon}(f^B)~s.t.~f(x')\neq f^B(x)}-1_{f(x)\neq f^B(x)})d\tilde{\mathbb{P}}(x)\\
   &=\frac{1}{K}\int_{\tilde{\mathcal{X}}}(1_{ \exists x' \in S_{x,\epsilon}(f^B)~s.t.~f(x')\neq f^B(x)})d{\mathbb{P}}(x)~~~\text{by}~f\in\mathcal{A}\\
   &\geq\frac{1}{K}\int_A(1_{ \exists x' \in S_{x,\epsilon}(f^B)~s.t.~f(x')\neq f^B(x)})d{\mathbb{P}}(x)= \frac{\mathbb{P}(A)}{K}>0.
\end{align*}
    Therefore, $\alpha>0$. Note that for any $f\in\mathcal{A}_\epsilon$, the second term in (\ref{eq:second_part}) is zero by the definition of $\mathcal{A}_\epsilon$. Therefore, first result of the theorem is proved. Furthermore, for $\epsilon$ such that $B(\tilde{\mathcal{X}},\epsilon)\supset {\mathcal{X}}$, $\mathcal{A}_\epsilon=\{f^B\}$. Therefore, when $B(\tilde{\mathcal{X}},\epsilon)\supset {\mathcal{X}}$, $f^B$ is the unique minimizer of $\tilde{\mathcal{R}}_{\rm rob}^s(f)$. \end{proof}

\begin{proof}[\rm \bf Proof of Corollary \ref{co:1}]
Because $A_\epsilon\subset A_0$, we get the result by the inequality 
\[\mathcal{R}_{\rm std}(f^*)\leq \max_{f\in A_\epsilon} \mathcal{R}_{\rm std}(f)\leq \max_{f\in A_0} \mathcal{R}_{\rm std}(f).\]
\end{proof}

\section{Omitted derivations}

\subsection{Example 1} \label{app:unif}
It is plain that the Bays classifier is $f^B(x)=sign(x_1-0.5)$ because with this classifier, $\mathcal{R}_{\rm{std}}(f^B)=0$ and obviously, other classifiers cannot achieve this zero risk. For the robust risk of the Bayes rule, the venerable area is $(0.5-\epsilon,0.5+\epsilon)\times (0,1)$
\begin{align*}
    \mathcal{R}_{\rm{rob}}(f^B)&= \mathcal{R}_{\rm{rob}}(f^B|Y=0)(1-p) + \mathcal{R}_{\rm{rob}}(f^B|Y=1)p\\
    &=\frac{\epsilon}{1/2}(1-p) + \frac{\epsilon}{1/2}p \\
    &= 2\epsilon.
\end{align*}

The robust model is $f_{\rm{rob}}(x)=sign(x_1-(0.5-\epsilon))$. Now, because the venerable area is only $(0.5-2\epsilon,0.5)\times (0,1),$
\begin{align*}
    \mathcal{R}_{\rm{rob}}(f_{\rm{rob}})&= \mathcal{R}_{\rm{rob}}(f_{\rm{rob}}|Y=0)(1-p) + \mathcal{R}_{\rm{rob}}(f_{\rm{rob}}|Y=1)p\\
    &=\frac{2\epsilon}{1/2}(1-p) + 0p \\
    &= (1-p)(4\epsilon).
\end{align*}
This is the possible minimum robust risk a model can achieve. For this model, the standard risk is
\begin{align*}
    \mathcal{R}_{\rm{std}}(f_{\rm{rob}})&= \mathcal{R}_{\rm{std}}(f_{\rm{rob}}|Y=0)(1-p) + \mathcal{R}_{\rm{std}}(f_{\rm{rob}}|Y=1)p\\
    &=\frac{\epsilon}{1/2}(1-p) + 0p \\
    &= (1-p)(2\epsilon).
\end{align*}
\subsection{Example 2} \label{app:cheese}

We  provide the derivations on the standard and adversarial risks. We first calculate the three classes of functions which minimize natural, adversarial robust, and sensibly robust risk respectively, w.r.t $\tilde{\mathbb{P}}_{X,Y}$. Then for each class, we consider the worst-case function from each class, in that the function maximizes the \emph{standard} risk w.r.t ${\mathbb{P}}_{X,Y}$. Likewise, for each class, we consider the worst-case function from each class, in that the function maximizes the \emph{adversarial robust} risk w.r.t ${\mathbb{P}}_{X,Y}$. 

First, the minimizers of each risk w.r.t. $\tilde{\mathbb{P}}_{X,Y}$ are as following.

1) Let $\tilde{\mathcal{F}}_B$ be a set of naturally optimal functions w.r.t $\tilde{\mathbb{P}}_{X,Y}$:
\[\tilde{\mathcal{F}}_B=\{f\in\mathcal{F}|f(x)= sign(x_1-0.5) \text{~for~} (x_1,x_2)\in \tilde{\mathcal{X}} \} .\]

2) Let $\tilde{\mathcal{F}}^s_{\rm rob}$ be a set of sensibly  optimal functions w.r.t $\tilde{\mathbb{P}}_{X,Y}$:
\[\tilde{\mathcal{F}}^s_{\rm rob}=\{f\in\mathcal{F}|f(x)= sign(x_1-0.5) \text{~for~} (x_1,x_2)\in B(\tilde{\mathcal{X}},\epsilon) \}. \]  

3) Let $\tilde{\mathcal{F}}_{\rm rob}$ be a set of robustly optimal functions w.r.t $\tilde{\mathbb{P}}_{X,Y}$:
\[\tilde{\mathcal{F}}_{\rm rob}=\{f\in\mathcal{F}|f(x)=sign(x_1-0.5+\epsilon) \text{~for~} (x_1,x_2)\in B(\tilde{\mathcal{X}},\epsilon) \}. \]

Second, we consider the worst-case standard risk w.r.t. ${\mathbb{P}}_{X,Y}$ for each class above.

1) $\max_{f\in \tilde{\mathcal{F}}_B}\mathcal{R}_{\rm std}(f)$: Although $p\neq 0.5$, due to the symmetry of the shape of $\tilde{\mathcal{X}}$, $\max_{f\in \tilde{\mathcal{F}}_B}\mathcal{R}_{\rm std}(f)$ is the area on $\mathcal{X}\setminus\tilde{\mathcal{X}}$, the area outside the small nine squares.

2) $\max_{f\in \tilde{\mathcal{F}}^s_{\rm rob}}\mathcal{R}_{\rm std}(f)$: For the same reason above, $\max_{f\in \tilde{\mathcal{F}}^s_{\rm rob}}\mathcal{R}_{\rm std}(f)$ is the area on $\mathcal{X}\setminus B(\tilde{\mathcal{X}},\epsilon)$, the area outside the small nine squares extended by $\epsilon$. 

3) $\max_{f\in \tilde{\mathcal{F}}_{\rm rob}}\mathcal{R}_{\rm std}(f)$: When $\epsilon <\alpha/4$, we consider the deviated line on $B(\tilde{\mathcal{X}},\epsilon)$, and regard the model outside $ B(\tilde{\mathcal{X}},\epsilon)$ as incorrect. The risk is calculated easily by using the fact that the risk of any $f \in \tilde{\mathcal{F}}_{\rm rob}$ on is $B(\tilde{\mathcal{X}},\epsilon)$ is $3\times\min(a+2\epsilon, 2a)\times(1-p)\times \min(2\epsilon, 0.5)/0.5$.

\subsection{Loss derivation} \label{app:loss_dev}
We begin from the definition of $\tilde{x}^s$ in (\ref{eq:sense_def2}). Then, from the definition of $\ell^{s}(f,x,y)$ in (\ref{eq:sense_loss2}),
\begin{align*}
    \ell^{s}(f,x,y)&= \ell(f(\tilde{x}^s),y) \\
    & = \ell(f(\tilde{x}^s),y)1_{A_f}+\ell(f(\tilde{x}^s),y)1_{B_f}+\ell(f(\tilde{x}^s),y)1_{C_f}\\
    & = \ell(f({x}),y)1_{A_f}+\ell(f(\tilde{x}^s),y)1_{B_f}+\ell(f(\tilde{x}),y)1_{C_f}\\
     & = \ell(f({x}),y)1_{A_f}+\log\frac{1}{c}1_{B_f}+\ell(f(\tilde{x}),y)1_{C_f}.
\end{align*}
The last equation is due to (\ref{equ:middle}) and the definition of $\tilde{x}^s$. Now, in the derivation of (\ref{eq:id_1}), we can see that
\[\ell(f({x}),y)1_{A_f}+\log\frac{1}{c}1_{B_f}+\ell(f(\tilde{x}),y)1_{C_f} =\ell^t_{\rm nat}(\hat p_y(x))+\ell^t_{\rm rob}(\hat p_y(\tilde{x})).\] Just for completeness, we show why we can see this. 
\begin{align*}
    For~x\in A_f,~ \ell(f({x}),y)&= -\log \hat{p}_y(x)=(-\log \hat{p}_y(x)-\log \frac{1}{c})+\log \frac{1}{c} \\&=\max\Big(0, -\log \hat p_y(x)-\log\frac{1}{c}\Big) + \min\Big(-\log \hat p_y(\tilde{x}),\log\frac{1}{c} \Big)\\& =\ell^t_{\rm nat}(\hat p_y(x)) + \ell^t_{\rm rob}(\hat p_y(\tilde{x})).\\
    For~x\in B_f,~ \log\frac{1}{c}&= 0+\log\frac{1}{c} = \max\Big(0, -\log \hat p_y(x)-\log\frac{1}{c}\Big) + \min\Big(-\log \hat p_y(\tilde{x}),\log\frac{1}{c} \Big) \\&= \ell^t_{\rm nat}(\hat p_y(x)) + \ell^t_{\rm rob}(\hat p_y(\tilde{x})).\\
    For~x\in C_f,~ \ell(f(\tilde{x}),y)&=-\log \hat p_y(\tilde{x}) = \max\Big(0, -\log \hat p_y(x)-\log\frac{1}{c}\Big) + \min\Big(-\log \hat p_y(\tilde{x}),\log\frac{1}{c} \Big) \\&= \ell^t_{\rm nat}(\hat p_y(x)) + \ell^t_{\rm rob}(\hat p_y(\tilde{x})).
\end{align*}
Therefore, we can conclude
\[  \ell^{s}(f,x,y) =\ell^t_{\rm nat}(\hat p_y(x))+\ell^t_{\rm rob}(\hat p_y(\tilde{x})) .\]

\section{The Three Clusters dataset experiment}
\subsection{Theoretical derivations}\label{app:three}

We provide the calculation of the standard and adversarial risks.
It is easy to see that the Bayes rule of this case is 
\begin{equation}\label{eq:bayese}
f^B(x)=
\begin{cases} 
+1,~~~~~~\text{if }p\exp{-\frac{\|x-\mu_A\|^2}{2\sigma^2}}+(1-p)\exp{-\frac{\|x-\mu_B\|^2}{2\sigma^2}}>\exp{-\frac{\|x-\mu_C\|^2}{2\sigma^2}}\\
-1,~~~~~~\text{otherwise}.\end{cases}
\end{equation}
It is straightforward to show that the standard risk of $f$ is 
\begin{align*}
    {\cal R}_{\rm std}(f_{\rm std})&=\mathbb{P}(f_{\rm std}(X)\neq y)\\&=\frac{\mathbb{P}(sign(X_1)>0|y=-1)}{2}+\frac{\mathbb{P}(sign(X_1)<0|y=1)}{2}\\&=\frac{\mathbb{P}(sign(N(-1,\sigma^2))>0)}{2}+\frac{\mathbb{P}(sign(N(1,\sigma^2)<0))}{2}\\&=\Phi(-\frac{1}{\sigma^2}).
\end{align*}
Therefore, as $\sigma^2\rightarrow 0$, the standard error ${\cal R}_{\rm{std}}(f_{\rm std})\rightarrow 0$ as well. However, the adversarial risk is
\begin{align*}
    {\cal R}_{\rm{rob}}(f_{\rm{std}})&=\mathbb{P}(f_{\rm std}(\tilde{X)}\neq y)\\&=\mathbb{P}(sign(X_1+\gamma\sigma)\neq y)\\&=\frac{\mathbb{P}(sign(N(-1,\sigma^2)+\gamma\sigma)>0)}{2}+\frac{\mathbb{P}(sign(N(1,\sigma^2)-\gamma\sigma)<0))}{2}\\&=\Phi(\gamma-\frac{1}{\sigma^2}).
\end{align*}

\section{Comparison to related work}\label{app:comp}
Here we present a methodological comparison of SENSE-AT to other related AT methods. TRADES \citep{zhang2019theoretically} is designed to trade adversarial robustness off against accuracy. 
The objective function of TRADES is 
    \begin{equation}\label{eq:trade}
    \mathcal{L}(f)=\E\big[\ell(Y, f(X)) + \lambda~\max_{\|X'-X\|_p\le \epsilon}\ell(f(X),f(X'))\big], 
    \end{equation}
    where the second term can be seen as a smoothness regularizer and $\lambda$ is the smoothing parameter. The possible interpretation of TRADES could be a replacement of the true label $Y$ by a softmax label $f(X)$ on its natural example as the prediction target on an adversarial example during training. This formulation shows several key differences between TRADES and SENSE-AT. First, TRADES controls the trade-off by controlling the weight between the natural and robust loss while SENSE-AT directly controls the lower bound on the prediction probability. Second, TRADES is trade-off between a standard model ($\lambda=0$) and a constant function ($\lambda=\infty$) \citep{zhang2019theoretically}. On the other hand, SENSE-AT focuses on the trade-off between a standard model and a robust model which is trained with full $\epsilon$ adversarial perturbations on the correctly classified natural examples. Third, most importantly, TRADES uniformly applies $\epsilon$ perturbation ball, whereas SENSE-AT applies adaptive perturbation sets for each input. If the model capacity is not enough to achieve the smoothness around an observation, (\ref{eq:trade}) can potentially lead to a less confident natural prediction.
Another related method is called MART \citep{Wang2020Improving}, short for Misclassification Aware Regularization, which can be seen as a variation of TRADES that takes misclassified examples into account. However, MART also does not consider adaptive perturbation sets either. 

 
\cite{balaji2019instance} proposed a method called IAAT, short for Instance Adaptive Adversarial Training. They noticed the tension between robustness and natural accuracy that arises when the perturbation set of different classes overlaps each other. To prevent this, \cite{balaji2019instance} adapted $\epsilon_i$ for each input $x_i$ and pursue adversarial robustness on this $\epsilon_i$-ball. Each $\epsilon_i\leq \epsilon$ is an optimized discretized value where the projected gradient descent (PGD) attacks on the $\epsilon_i$-neighborhood of $x_i$ cannot lead to a missclassification. SENSE-AT is similar with IAAT in that a strongly misleading adversarial example is not observed during training. On the other hand, we use a single, fixed $\epsilon$ and instead directly adapt the perturbation set, not necessarily a ball, which is computationally much more efficient because we do not optimize $\epsilon$ for each individual input. Further, a strong incentive for the adapted perturbation sets to become the full $\epsilon$-balls is a unique property of the implicit loss truncation of SENSE-AT. If we do not replace of the \emph{incorrect prediction} condition by the \emph{confident prediction} condition in the re-defined sensible adversarial example, IAAT can be seen as a special case of SENSE-AT. 

\cite{Ding2020MMA} proposed a method called MMA, short for Max-Margin Adversarial training. 
The goal of MMA is, with $\ell$ as the cross-entropy loss, to minimize
\[R_{\rm mma}(f)=\E\Big\{\max(0, d_{\rm max}-d_f(x,y))1_S+\lambda \ell(f(x),y))1_{S^c} \Big\},\] where $S={\{(x,y):f(x)= y\text{ and }d_f(x,y)>0\}}$, $d_f(x,y)$ is the input margin, and $\lambda$ is the smoothing parameter. The threshold $d_{max}$ is the hyperparameter which is generally set to be greater than $\epsilon$, because otherwise the model's performance on attacks with norm $\epsilon$ becomes irrelevant to the objective. 
%
It is reported that finding a proper $d_{max}$ in practice is difficult \citep{Ding2020MMA}. 
%
%
We note that the input margin does not necessarily lead to a confident prediction on the perturbed neighborhood of the input. However, as long as a natural input is correctly classified and its input margin is greater than $d_{max}$, its adversarial examples are ignored no matter how uncertain the prediction on them can be.
    Only for the correctly classified inputs with the smaller margin than $d_{max}$, robustness is pursued against attacks of prediction confidency up to 0.5. 

\begin{supplement}
\stitle{Code and data}
\sdescription{We provide a Python package and trained models to reproduce the results shown in our experiments, and also R code for the Three Clusters dataset.}
\end{supplement}

\bibliographystyle{chicago}
\bibliography{refer}

\end{document}